\documentclass[11pt,a4paper]{article}

\usepackage[a4paper,margin=1in,includehead,headheight=60pt,headsep=12pt]{geometry}

\newcommand*{\doi}[1]{%
   \href{https://doi.org/#1}{doi:~{\small\nolinkurl{#1}}}}
\usepackage{graphicx}
\usepackage{fancyhdr}
\usepackage{tikz}
\usepackage[hidelinks]{hyperref}
\usepackage[numbers]{natbib}
\usepackage{pgffor} 

\usepackage{longtable}
\fancypagestyle{firstpage}{
  \fancyhf{}

}

\usepackage{subcaption}  
\usepackage{caption}     

\usepackage{tikz}
\usepackage{pgfplots}
\pgfplotsset{compat=1.18}
\usepackage{xspace}
\usepackage{forest}
\definecolor{folderbg}{RGB}{124,166,198}
\definecolor{folderborder}{RGB}{110,144,169}
\def\Size{4pt}
\tikzset{
  folder/.pic={
    \filldraw[draw=folderborder,top color=folderbg!50,bottom color=folderbg]
      (-1.05*\Size,0.2\Size+5pt) rectangle ++(.75*\Size,-0.2\Size-5pt);  
    \filldraw[draw=folderborder,top color=folderbg!50,bottom color=folderbg]
      (-1.15*\Size,-\Size) rectangle (1.15*\Size,\Size);
  }
}
\newcommand{\gridfull}{Universal grid Representation of Showers\xspace}
\newcommand{\grid}{Universal grid Representation\xspace}

\title{\dataset dataset: \datasetfull}
\author{Peter McKeown, Piyush Raikwar, and Anna Zaborowska\\European Organisation for Nuclear Research (CERN), Switzerland}
\date{\today}
\newcommand{\dataset}{LEMURS\xspace}
\newcommand{\datasetfull}{Large-scale multi-detector ElectroMagnetic Universal Representation of Showers\xspace}
\usepackage{graphicx}

\begin{document}
\maketitle
\begin{tikzpicture}[remember picture,overlay]
  \node[anchor=north east, xshift=-1.2cm, yshift=-1.2cm]
    at (current page.north east) {\includegraphics[height=3cm]{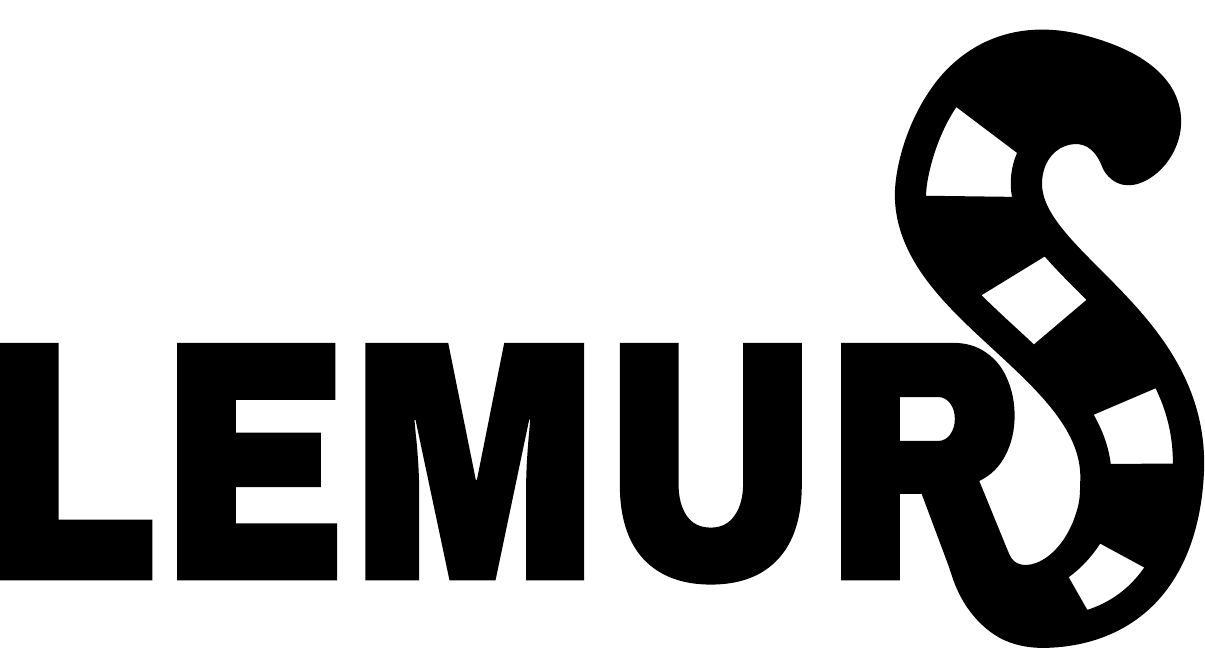}};
\end{tikzpicture}

\begin{abstract}
We present \dataset: an extensive dataset of simulated calorimeter showers designed to support the development and benchmarking of fast simulation methods in high energy physics, most notably providing a step towards the development of foundation models. This new dataset is more robust than the well-established CaloChallenge dataset 2, featuring substantially greater statistics, a wider range of incident angles in the detector, and most crucially multiple detector geometries with more realistic calorimeters. \dataset's scale and diversity make it particularly suitable for development of generalisable models and has been used in the development of CaloDiT-2, a pre-trained model released in the community standard simulation toolkit \textsc{Geant4}. All data and code for generation and analysis are openly accessible, facilitating reproducibility and reuse across the community.
\end{abstract}

\section{Background \& Summary}

High-energy physics experiments rely heavily on detailed simulations of particle interactions with detectors. One of the most computationally expensive components of these simulations is the interaction with the calorimeter, a sub-detector that measures the energy of particles by absorbing them (the incident particle initiates a shower or a cascade). Traditional (full) detector simulation, performed with \textsc{Geant4}~\cite{Geant4}, offers high fidelity but requires significant computational resources and time -- especially for the increasing number of collisions at the experiments which will be operated at the High Luminosity Large Hadron Collider~\cite{Software:2815292,CERN-LHCC-2022-005}.

To address this challenge, fast calorimeter simulation methods have been developed, starting with~\cite{GFlash,Paganini:2017dwg,ATLAS_VAE_GAN2018,Atlfast3,Buhmann_2021}. In particular, machine learning-based generative models have emerged as promising tools. They learn to replicate the complex distributions of calorimeter responses from training data, enabling significant speed-up while attempting to maintain sufficient accuracy for many physics analyses.

Most models are developed within the scope of a particular detector, often not offering the possibility for external contributors to collaborate and use the shower datasets. A greatly appreciated exception was the activity proposed within the CaloChallenge~\cite{Krause:2024avx}, where three datasets were published, with the first one (dataset\,1) coming from the ATLAS public dataset (for electromagnetic (EM) and hadronic showers~\cite{michele_faucci_giannelli_2023_8099322_ds1}), and the latter two coming from the \textsc{Geant4} Par04 example~\cite{Par04}, both of which feature a higher granularity: dataset\,2~\cite{faucci_giannelli_2022_6366271_ds2} with 
6'480 voxels and dataset\,3~\cite{faucci_giannelli_2022_6366324_ds3} with 
40'500 voxels. For comparison, dataset\,1 has a granularity of 368 for EM and 533 for pion showers. CaloChallenge has resulted in many contributions and allowed a first common benchmark of fast calorimeter simulation to be performed for a large number of models. Its open datasets and validation metrics can be used to compare any new model with the previously submitted ML models.
The \dataset dataset has a similar structure and the same dimensionality as dataset 2 of CaloChallenge, providing sufficiently large granularity to challenge the model development, targeting a level appropriate for most of the current experiments. The representation is targeted at EM showers. The file structure is similar, with two angles describing the incident particle being added (only energy was present in CaloChallenge), and the shape of the individual showers changed from flat $z \times \varphi \times r$ (following the ATLAS-inspired dataset~\cite{michele_faucci_giannelli_2023_8099322_ds1}) to 3D $r \times \varphi \times z$. The energy scale is also not calibrated as it was done in CaloChallenge, more details on sampling fraction can be found in Sec.~\ref{sec:sampling_fraction} and Table~\ref{tab:sf}.

However, a fast simulation model that can be applied in a realistic detector must address several challenges that were not a topic of the CaloChallenge. Starting with the distribution of particles in the detector, the CaloChallenge focused on varying the incident energy of particles, with all particles incident in the same part of the detector. This concept originally comes from ATLAS, the only Large Hadron Collider experiment that uses ML models in production, where currently one model is trained for each of the small slices of the detector~\cite{Atlfast3}. This tackles the problem of the detector structure varying. Alternatively, an ML model can be also be conditioned on the incident angle during training -- the \dataset dataset enables this approach.

Secondly, CaloChallenge dataset 2 and 3 (not dataset 1), were produced with a simplistic demonstrator of a calorimeter, with layers of material arranged in a cylindrical structure. This provides a valuable dataset for benchmarks, but it lacks the geometrical complexities of a realistic detector, including discontinuities, transition regions, etc. The \dataset dataset includes this demonstrator with two different readout technology options, as well as additional detectors with more complex geometries, in order to study fast calorimeter simulation under more realistic conditions.

Finally, the inclusion of different detector geometries allows a step to be made towards foundation models in HEP. This dataset has been used to build CaloDiT-2, a pre-trained model that is distributed with Geant4~\cite{Geant4,Par04}, a standard toolkit used in HEP (and beyond) for particle transport simulation. Application of this model to any new detector is much quicker and requires much less data than training the model from scratch. The \dataset dataset opens a door towards studies in this direction, with possibilities to extend to tasks beyond simulation~\cite{CaloDiT}. An option to use a single detector dataset is of course also possible.

\section{Methods}
\subsection{Nomenclature}

The following description is a simplification to illustrate the main concepts and terminology that would enable an understanding of the dataset.

When a particle enters the detector, it travels through the detector and interacts with its materials, depositing energy and creating secondary particles. The energy that the detector measures and its placement (in space or time) can be called a \textbf{detector response}. This is measured in a certain structure, with the detector divided into sub-detectors, layers, cells, etc. which will be referred to as the \textbf{detector readout}. The smallest physical unit of readout is a \textbf{cell}. An \textbf{event} is a detector response to a certain input (\textbf{primary particles}), which is typically complex, involves many particles and corresponds to the modelling of a particle collision, but in this instance the focus is solely on single particle inputs. The particles that enter the sub-detector called the \textbf{calorimeter}, whose purpose is to measure the energy of incident particles by absorbing (stopping) them, create \textbf{cascades} of secondary particles called \textbf{showers}. If the incident particle is an electron e$^-$, positron e$^+$, or photon $\gamma$, almost exclusively electromagnetic interactions with matter occur, hence the showers they produce are called \textbf{electromagnetic (EM) showers}. If a hadron interacts with the calorimeter (e.g. a pion $\pi^\pm$ or proton p), it can interact via the strong interaction, as well as by electromagnetic interactions, forming a \textbf{hadronic shower}. Hadronic showers are significantly more complex and exhibit larger variations than the EM showers, and are not in scope of this dataset. In the \textbf{full simulation} performed with Geant4~\cite{Geant4}, each simulated particle (\textbf{MC particle} = Monte Carlo particle) traverses the detector in little \textbf{steps}, and at each one there could have been energy deposited and/or secondary particles created. Sum of energy deposits with a detector cell is made in the typical simulation (mimicking the reality), but for shower parameterisation studies, this sum is performed over a much smaller volume called a voxel and described in Sec.~\ref{sec:grid}.

\subsection{Simulation}

The prior dataset that was produced for CaloChallenge directly used the simulation application released with Geant4, called the extended example Par04~\cite{Par04}. It demonstrates how to produce a shower dataset, and how to use a model for inference. It also defines the representation of the shower, which was up to now given different names (e.g. ``Par04 scoring mesh'', ``CaloChallenge cylindrical data''), while we have decided to name it uniformly a \gridfull. It is described in detail in \ref{sec:grid}.

As the \dataset dataset features more than a single detector, and goes beyond the simplistic demonstrator detectors with up to 2 materials, a different framework was used to simulate the data. For consistency, the demonstrator detectors were also simulated in this new library. It relies heavily on the key4hep software stack~\cite{Carceller:2025ydg}, a community choice for the software for future collider detectors. As the choice for the simulation framework within key4hep is DD4hep~\cite{Gaede:2020tui} (or more precisely the DDG4 component), a new library has been prepared to demonstrate how to use the fast simulation hooks of Geant4 within DDG4, called ddfastsim~\cite{ddfastsim,ddfastsim_presentation}. It implements all the necessary components and steering scripts to run fast simulation, and in particular it implements the \grid and the geometry description of the Par04 demonstrator detector. It also includes the scripts that were used to run a production on the HTCondor batch service~\cite{htcondor}. Finally, the translation of the output in the native format to the commonly used HDF5 file format is provided.

\subsection{Detectors and coordinate system}
\label{sec:detectors}

\begin{figure}[htbp]
  \centering
  \begin{subfigure}[b]{0.5\textwidth}
\includegraphics[width=\textwidth,page=1,trim={0.4cm 0.4cm 0.4cm 0.4cm},clip]{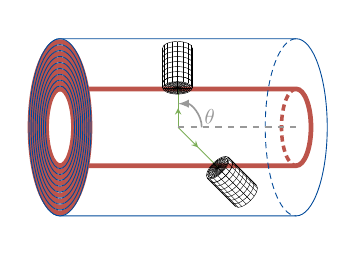}
    \caption{3D illustration of the detector, with $\theta$ angle definition.}
    \label{fig:theta}
  \end{subfigure}
  \hfill
  \begin{subfigure}[b]{0.4\textwidth}
~\includegraphics[width=\textwidth,page=2,trim={0.4cm 0.4cm 0.4cm 0.4cm},clip]{images/cylinder_detector-8.pdf}
    \caption{Transverse projection of the detector and the definition of the $\phi$ angle.}
    \label{fig:phi}
  \end{subfigure}
  \caption{Depiction of the demonstrator detector originating from the Par04 example. Additionally, an explanation of the coordinate system is provided. Incident showers enter the detector at different angles: $\theta$ describes the angle with respect to the beam (detector) axis, and $\phi$ defines the direction in the transverse plane of the detector. The incident direction defines the structure of the \grid voxelisation.}
  \label{fig:detector_par04}
\end{figure}

The \dataset dataset features 5 detectors, and each of them can be either used individually, or combined. For each detector, showers were simulated in the barrel region of the electromagnetic calorimeter (ECAL). The demonstrator detector, Par04, is composed of up to 2 materials, with alternating layers of passive and active material in a cylindrical barrel geometry. The geometry that is defined in the Par04 example of Geant4, and that was used to produced dataset 2 of the CaloChallenge, is called \emph{Par04\_SiW}, and it is composed of 90 layers of $1.4$\,mm tungsten absorbers and $0.3$\,mm silicon sensitive layers. A different set of materials is also present in \emph{Par04\_SciPb} detector, with 45 layers of $4.4$\,mm lead absorbers and $1.2$\,mm scintillators. The illustration of such setups is depicted in Fig.\ref{fig:detector_par04}. This picture also shows how the coordinate system is defined. All particles produced in simulation are produced in the centre of the detector, at $(x,y,z) =(0,0,0)$, and can travel in the direction defined by two angles: the polar angle $\theta$ that describes the inclination towards the beam axis (= the cylinder axis), Fig.~\ref{fig:theta}, and the azimuthal angle $\phi$ that describes the inclination in the transverse plane, defined with respect to the $x$ axis of the detector, Fig.~\ref{fig:phi}. 

To complement those two simplistic detectors, three other, more realistic detectors have been added: the Open Data Detector~\cite{odd}, a benchmark detector for algorithm development, and two detectors proposed for the electron-positron Future Circular Collider (FCC-ee): CLD~\cite{Bacchetta:2019fmz} and ALLEGRO~\cite{Mlynarikova:2025skz}. As the focus of this study is electromagnetic showers, meaning those which occur in the electromagnetic calorimeter, the tracking detectors and the beampipe have been removed from the simulation. This avoids any early photon conversions prior to them reaching the calorimeter volume. Only events with single incident photons are included in the dataset.

\begin{figure}[htbp]
    \centering
    \begin{minipage}{0.45\textwidth}
        \centering
        \includegraphics[width=1\textwidth]{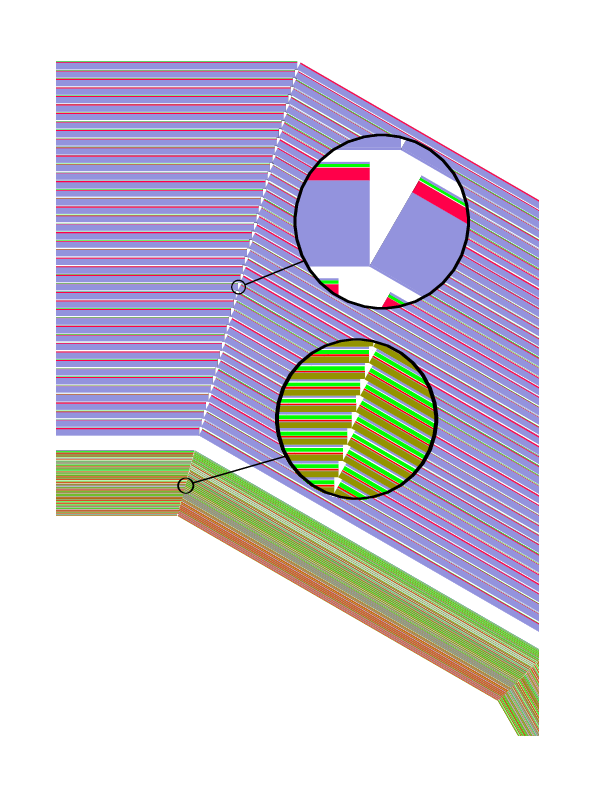}
        \caption{The complexity of the implementation of the calorimeter layers in simulation, with different materials being represented in different colours. Discontinuity between the modules of the polygon are shown in the zoom-in~\cite{Bacchetta:2019fmz}.}
        \label{fig:detector_cld_polygon}
    \end{minipage}\hfill
    \begin{minipage}{0.45\textwidth}
        \centering
        \includegraphics[width=1\textwidth]{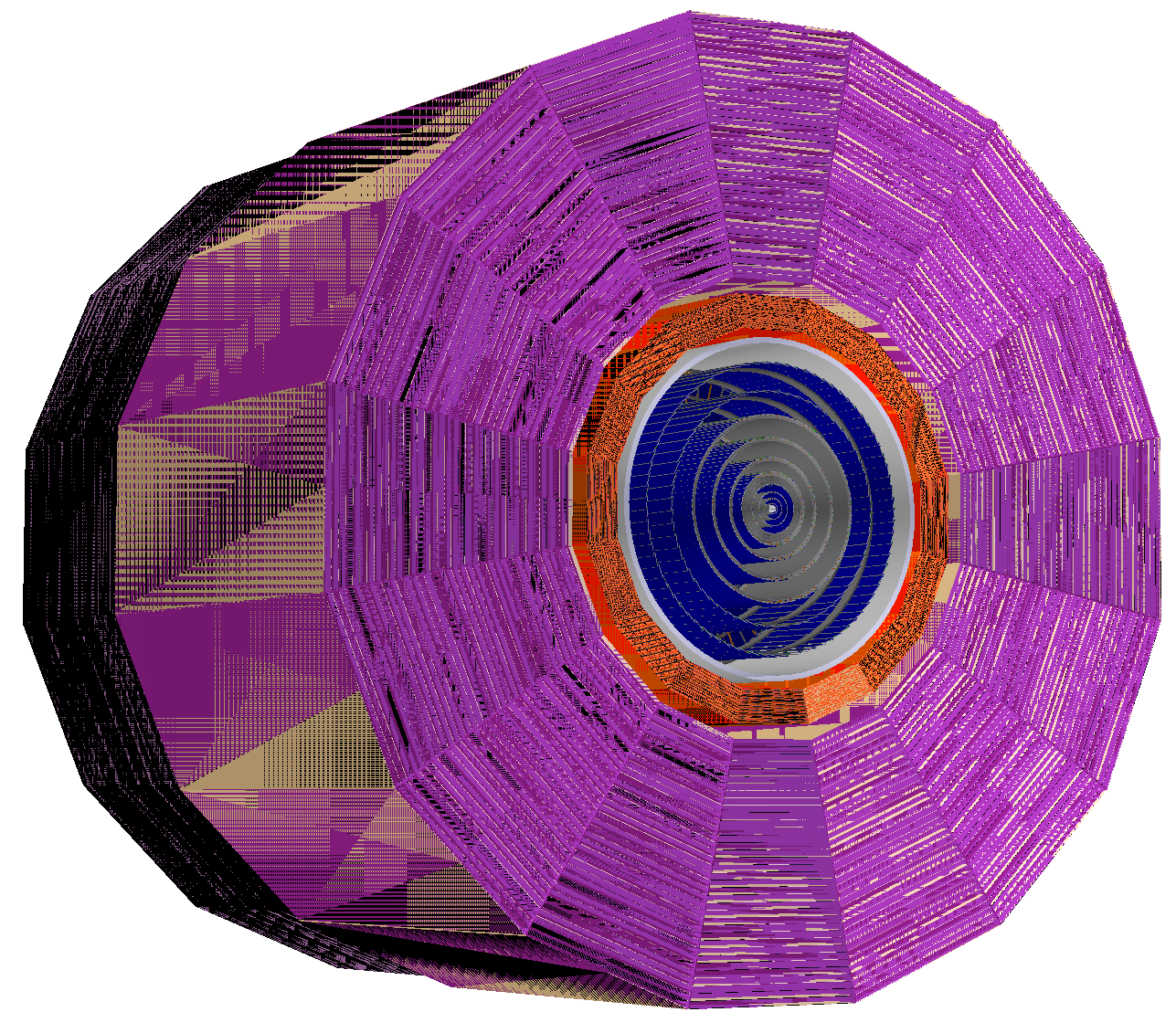}
        \caption{The Open Data Detector (ODD), with the EM calorimeter depicted in red. In the transverse plane, the individual layers placed in a hexadecagonal structure can be seen. The larger violet structure shows the hadronic calorimeter.}
        \label{fig:detector_odd}
    \end{minipage}
\end{figure}

\begin{figure}[htbp]
  \centering
  \begin{subfigure}[b]{0.45\textwidth}
    \includegraphics[width=\textwidth,]{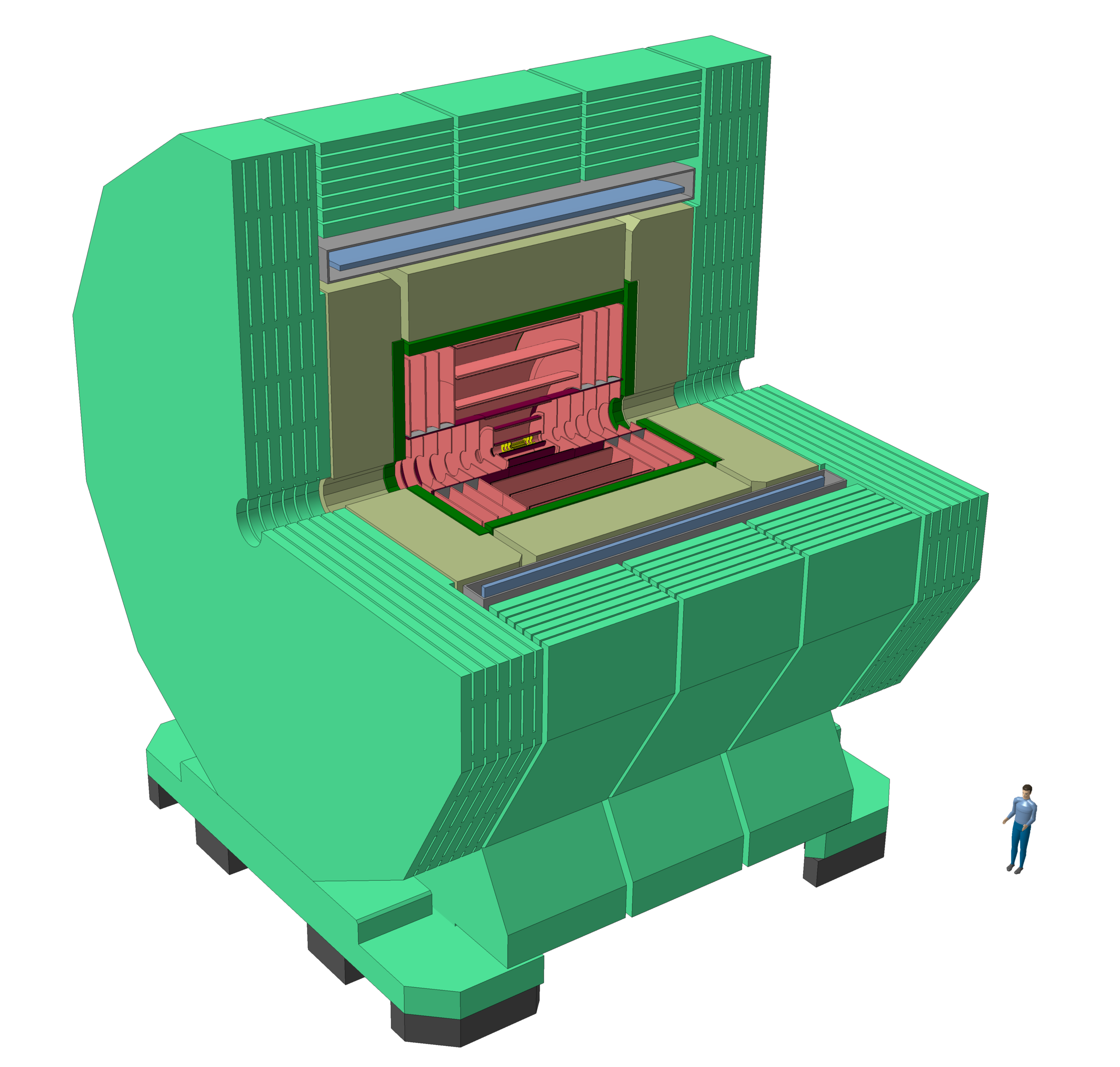}
    \caption{3D view of the detector, with one quarter removed.}
  \end{subfigure}
  \hfill
  \begin{subfigure}[b]{0.45\textwidth}
\includegraphics[width=\textwidth]{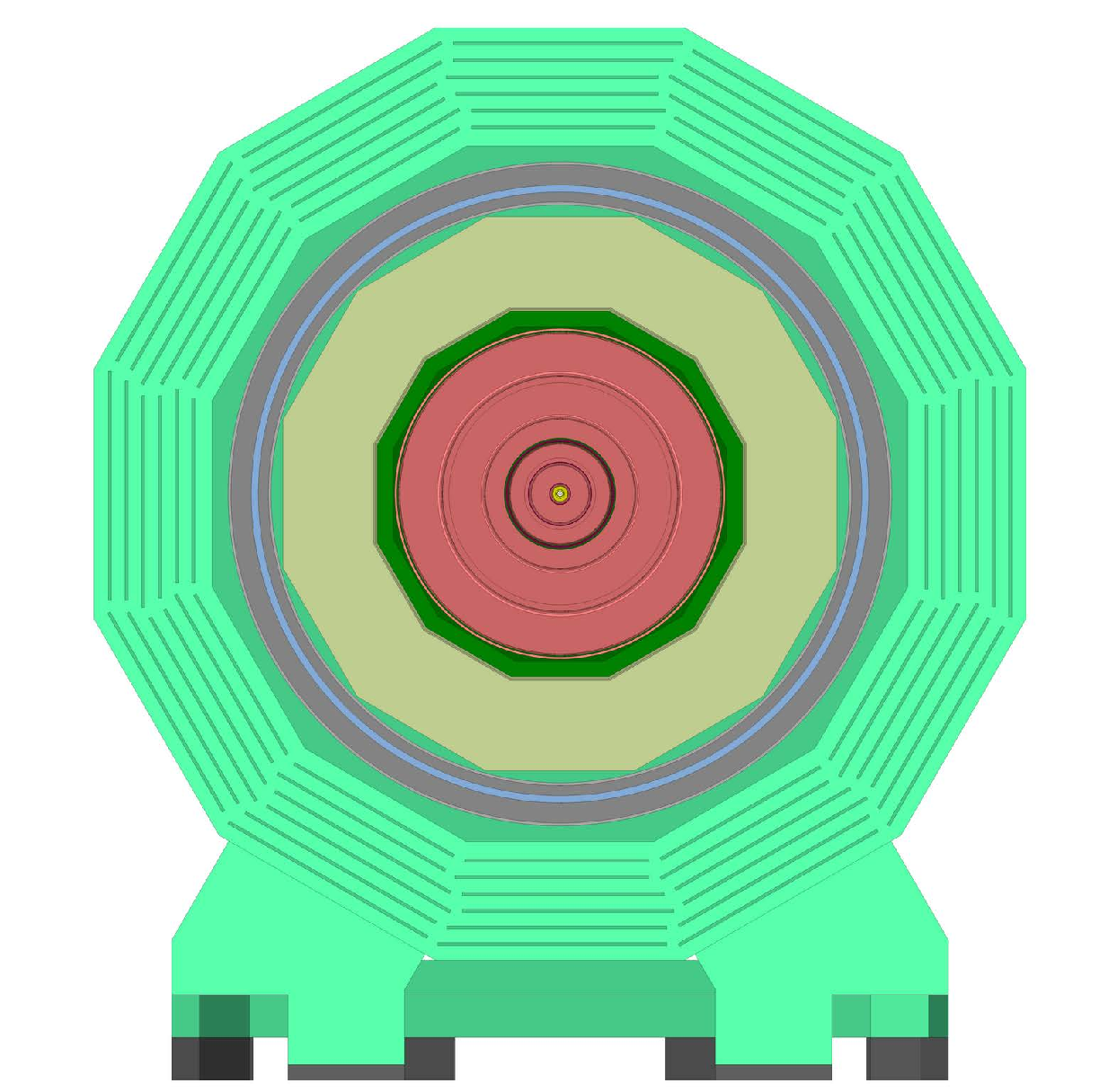}
    \caption{Transverse (XY) cross section.}
  \end{subfigure}
  \caption{The FCCee CLD detector, with the EM calorimeter depicted in dark green~\cite{Bacchetta:2019fmz}.}
  \label{fig:detector_cld}
\end{figure}

The Open Data Detector (\emph{ODD}, Fig.~\ref{fig:detector_odd}), as well as CLD (\emph{FCCee\_CLD}, Fig.~\ref{fig:detector_cld}) use sandwich layers in the calorimeters, with the barrel region of the ECAL featuring a polygonal cross-section in the transverse plane. This type of calorimeter structure is depicted in Fig.~\ref{fig:detector_cld_polygon} for CLD, with a zoom-in of regions of the detector which are particularly irregular. The EM calorimeter of FCC-ee CLD (version \texttt{FCCee\_o2\_v04} from k4geo~\cite{k4geo}) includes 40 sampling layers with $1.9$\,mm tungsten, $0.5$\,mm silicon, and other materials which have less influence on shower development ($2.65$\,mm of PCB, glue and air), placed in a dodecagonal (12-sided polygon) structure. The EM calorimeter of ODD was heavily inspired by CLD, with the main differences being the increased number of layers to 48, and a layer placement in a hexadecagonal (16-sided polygon) structure. 



\begin{figure}[htbp]
  \centering
  \begin{subfigure}[b]{0.45\textwidth}
    \includegraphics[width=\textwidth]{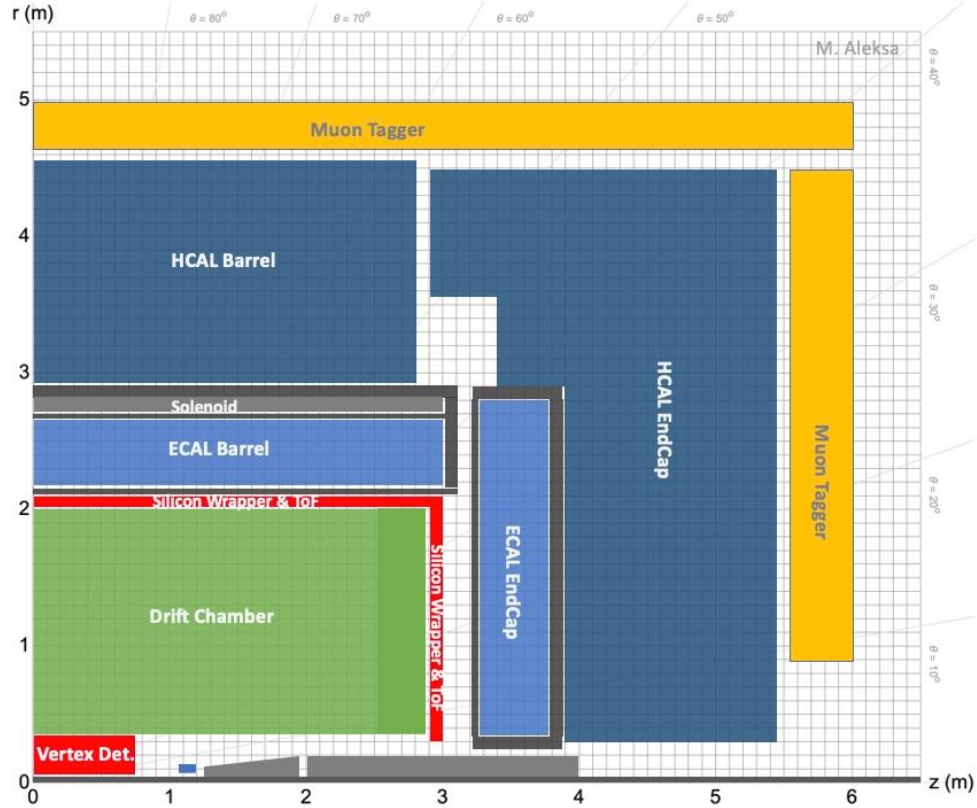}
    \caption{A quadrant of the detector. EM calorimeter is depicted with light blue colour.}
  \end{subfigure}
  \hfill
  \begin{subfigure}[b]{0.45\textwidth}
    \includegraphics[width=\textwidth]{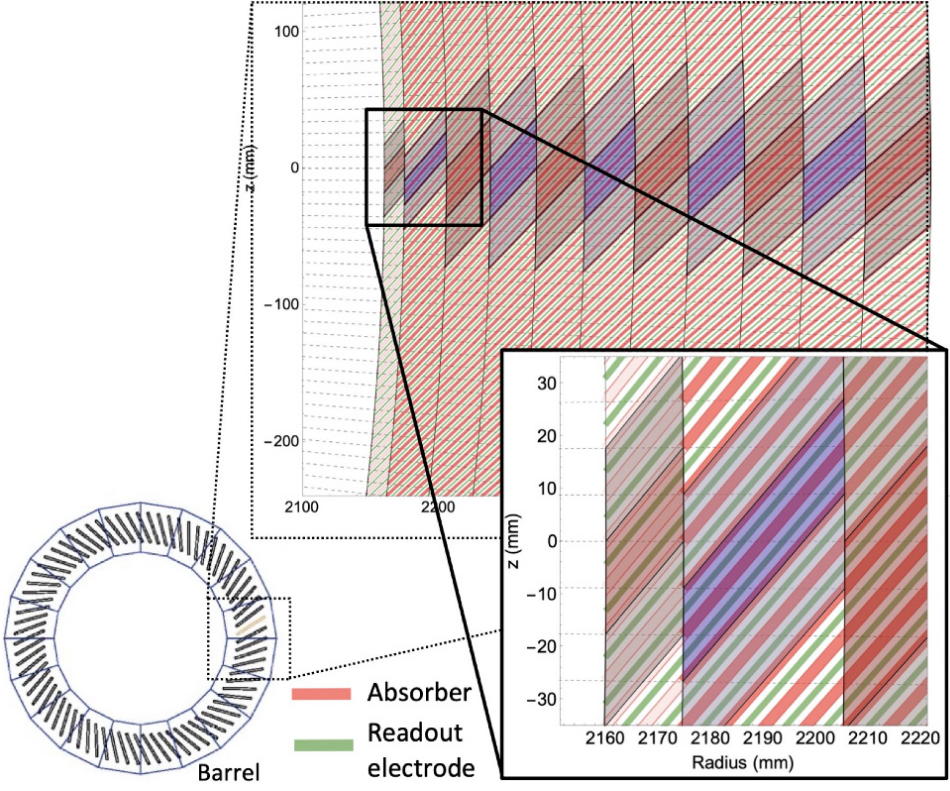}
    \caption{The EM calorimeter, with a cross section and zoom on the layers.}
  \end{subfigure}
  \caption{The FCCee ALLEGRO detector~\cite{Mlynarikova:2025skz}.}
    \label{fig:detector_allegro}
\end{figure}

The layout of the electromagnetic calorimeter of the ALLEGRO detector (\emph{FCCee\_ALLEGRO}, Fig.~\ref{fig:detector_allegro}, version \texttt{ALLEGRO\_o1\_v02} from k4geo~\cite{k4geo}) is not at all similar to the ones discussed so far. While there is no division into (polygon) modules in the azimuthal angle, there is a structure in the form of thousands (1.5k) of inclined modules with increasing thickness. These inclined models consist of active noble liquid accompanied by a lead absorber and readout electronics. The number of readout layers is much lower (12 layers), due to the manner in which the electronic signals are read out, with the individual cells being inclined with respect to the radial direction. This means that ALLEGRO is another realistic, detector geometry, that was included in the \dataset dataset.

\subsection{Particle shower representation: The \grid}
\label{sec:grid}

\begin{figure}[htbp]
\centering
    \begin{subfigure}{0.35\textwidth}
        \centering
         \includegraphics[width=\textwidth]{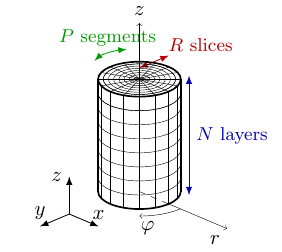}
    \caption{Cylinder used to record the energy of a shower. The incident particle direction defines the $z$ axis of the cylinder.~\cite{SALAMANI2023138079}}
        \label{fig:grid}
    \end{subfigure}\hfill
    \begin{subfigure}{0.6\textwidth}
        \centering
        \includegraphics[width=0.5\textwidth,trim={23.5cm 15cm 13cm 7cm},clip,angle=-90]{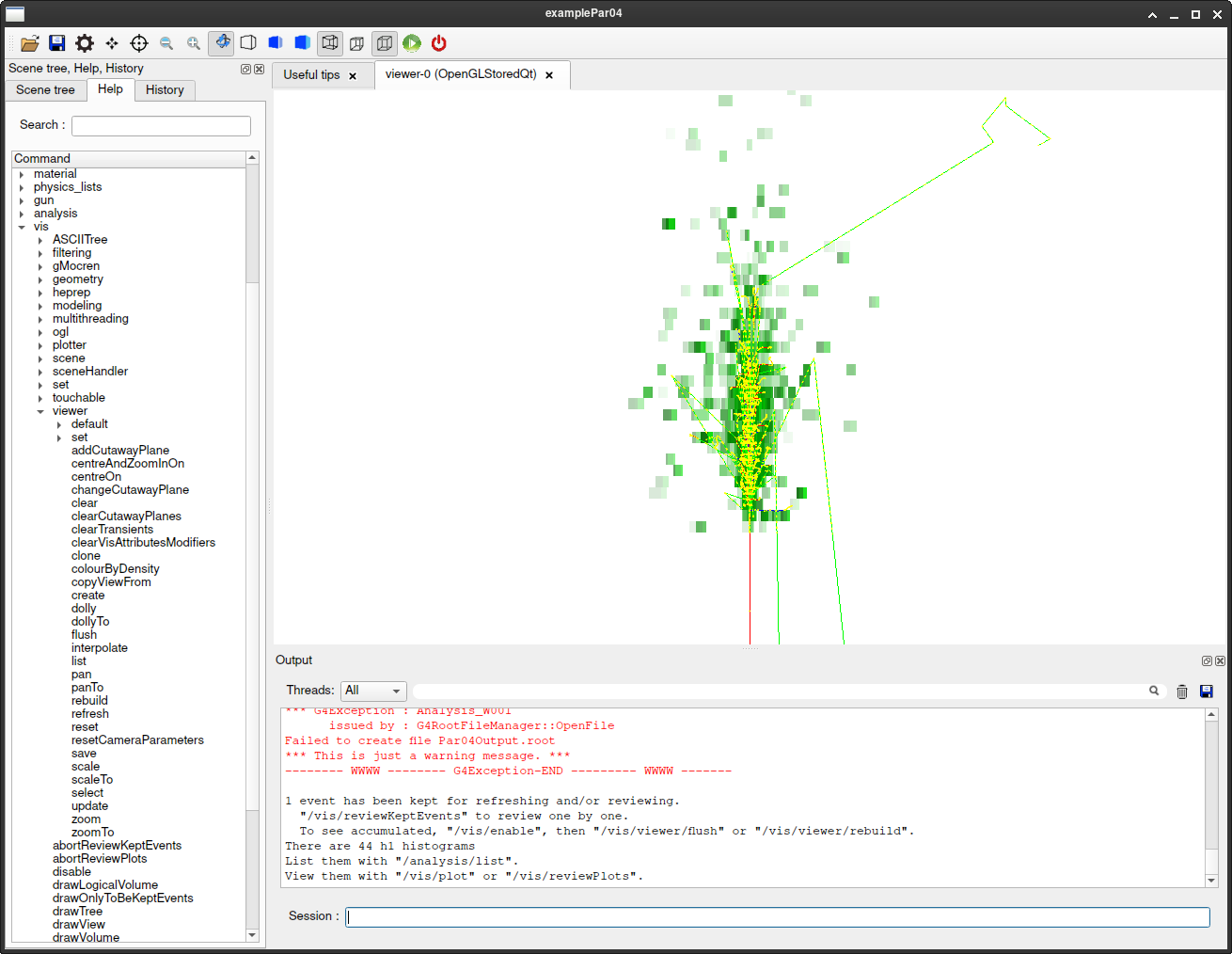}\begin{tikzpicture}
	       [overlay,thick, xshift=-8cm,yshift=-2.5cm,scale=3]
	       \draw[-latex] (0,0) -- node (Mid)  {} ++(0.5,0) node[above] {$z$};
 	          \draw[-latex] (0,0) -- ++(0,0.5) node[above] {$x$};
	       \end{tikzpicture}
    \caption{Projection along the axis of the shower development of a single $10$\,GeV EM shower. The intensity of the green voxels is linked to the amount of the deposited energy. The coloured lines correspond to the particles simulated within the shower in \textsc{Geant4}. Only high energy particles are left, as otherwise the voxels would be not visible. These visualisations were produced with the event display of the Par04 example.}
        \label{fig:shower_xz}
    \end{subfigure}
    \begin{subfigure}{0.45\textwidth}
        \centering
\includegraphics[width=\textwidth,trim={20cm 15cm 10cm 5cm},clip]{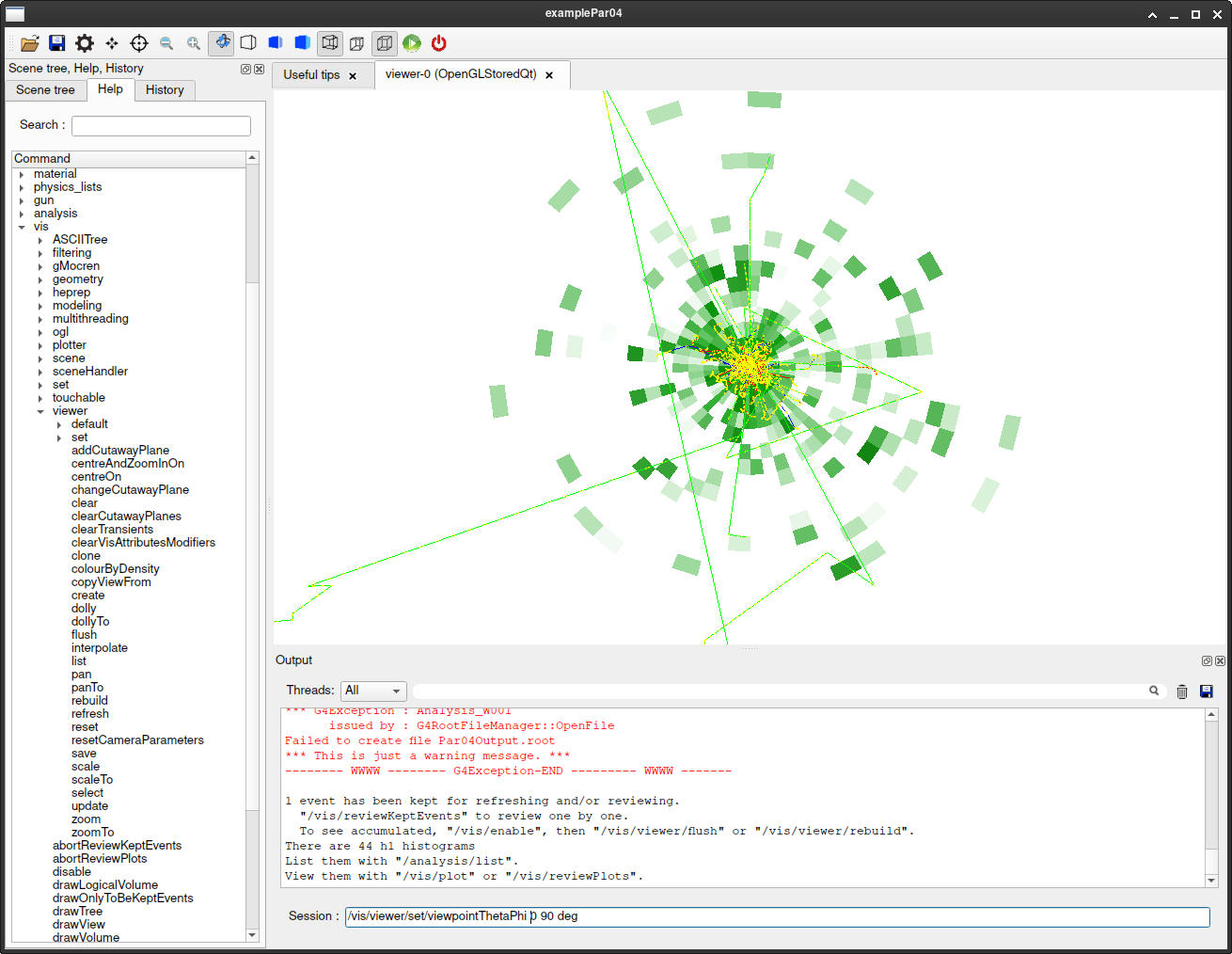}\begin{tikzpicture}
    	[overlay,thick,
    	xshift=-7cm,yshift=0.cm, scale=3]
    	\draw[-latex] (0,0) -- node (Mid)  {} ++(0.5,0) node[above] {$x$};
     	\draw[-latex] (0,0) -- ++(0,0.5) node[above] {$y$};
       \end{tikzpicture}
    \caption{Transverse projection of a single $10$\,GeV EM shower. Representation analogous to the longitudinal projection.}
        \label{fig:shower_xy}
    \end{subfigure}\hfill    \begin{subfigure}{0.5\textwidth}
        \centering
         \includegraphics[width=0.75\textwidth]{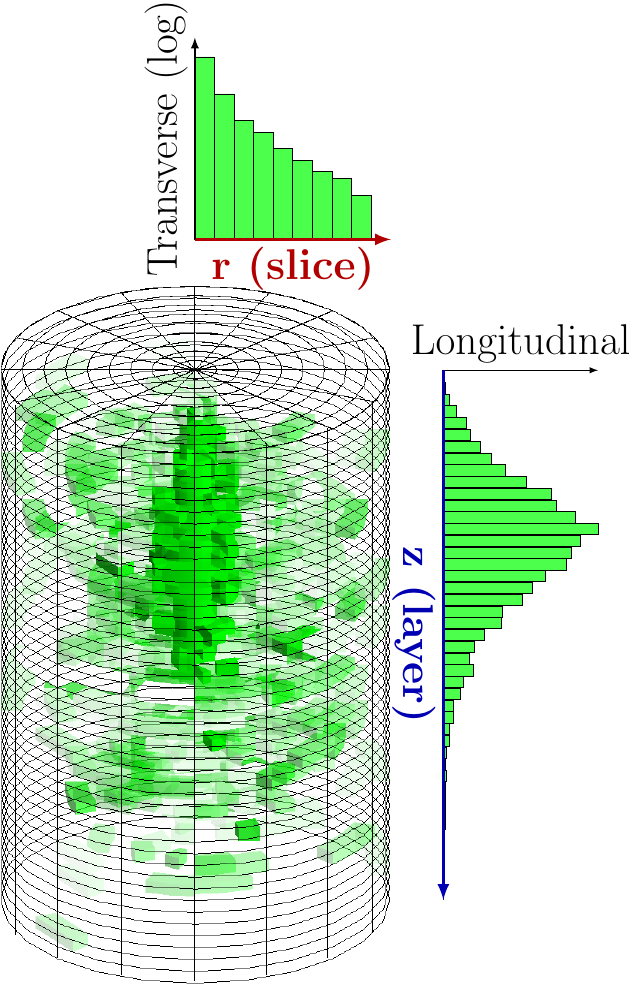}
    \caption{Illustration of a single shower and how profiles are constructed: transverse profile projects energy depositions on radial direction, while longitudinal profile is a projection on longitudinal direction.}
        \label{fig:grid_profiles}
    \end{subfigure}
    \caption{\grid: (a) its definition, (b) a transverse projection of an example shower, (c) a corresponding longitudinal projection.}
\end{figure}
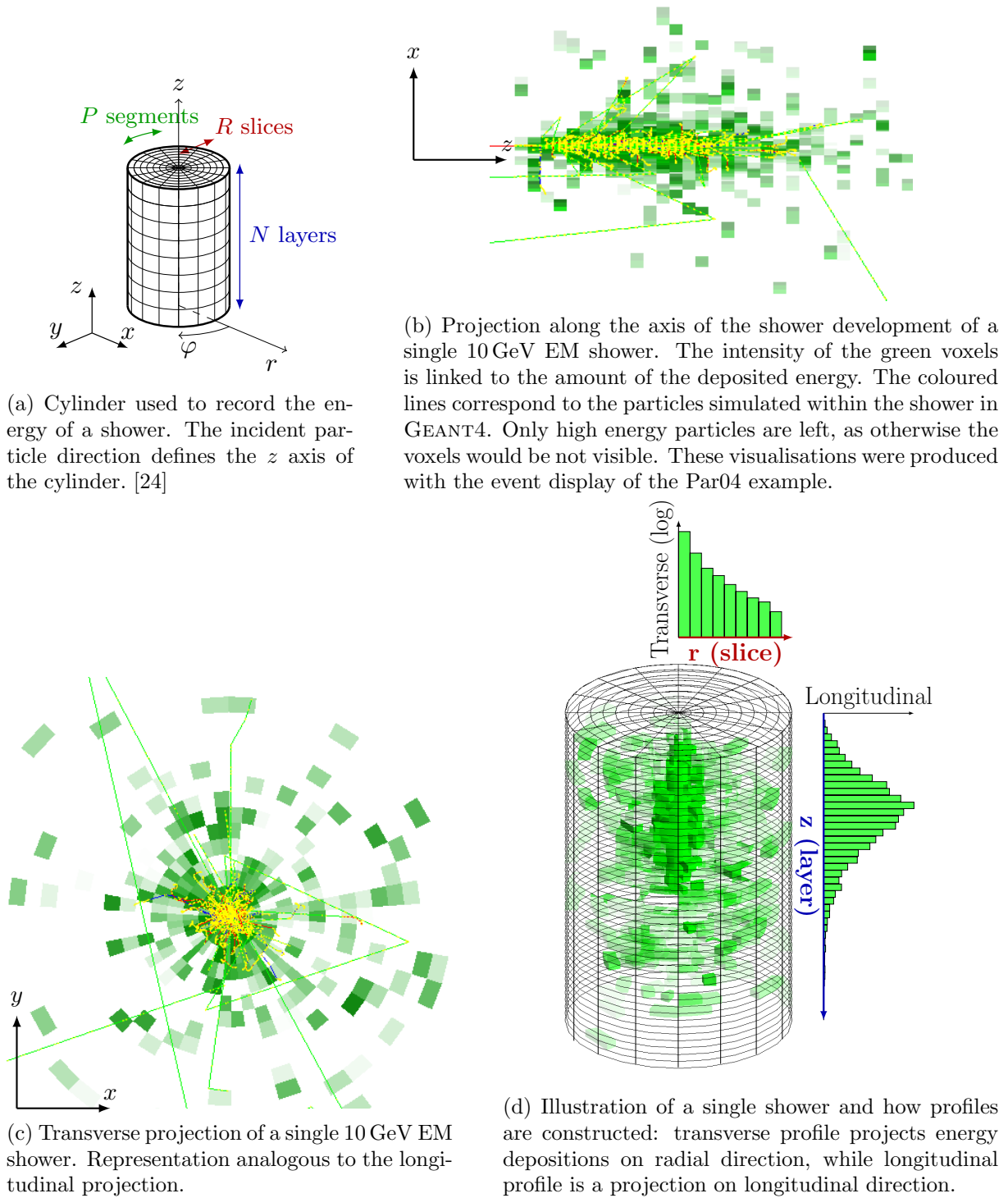

Incident photons, electrons, and positrons create cascades of particles (electromagnetic showers) once they start interacting in the dense material of calorimeters, as shown in Fig.~\ref{fig:cascade}. Showers are usually described in terms of their development along the initial particle direction (longitudinal), and in the transverse plane (or radially, as approximate symmetry in the azimuthal angle is expected). As energy deposits can be represented as 3D pictures, a choice of a cylindrical coordinate system seems natural. Such a cylinder needs to be placed at the entrance of the EM calorimeter, with the cylinder axis $z$ defined by the particle direction ($z=0$ at the entrance), and the centre in the transverse plane at the position that the particle enters the detector. They are depicted in Fig.~\ref{fig:detector_par04}. In order to record granular information about the showers, the cylinders are segmented in terms of number of layers (along $z$), number of radial slices (along $r$) and number of segments (in $\varphi$), as presented in Fig.~\ref{fig:grid}. An individual cell of such a cylindrical grid will be referred to as voxel. Several energy deposits may be merged into one voxel.

This representation of showers within the detectors is called a \grid. It is independent of the readout structure of the detector, and a placement back into the detector is needed after the output from any fast simulation model is produced. However, there are clear benefits of such an approach. Firstly, showers can be represented in the same structure, no matter in which detector and in which part of the detector they appear. Secondly, this allows the shower to be recorded at a granularity higher than that of the calorimeter readout. Finally, they allow models to be transferred more easily between different detectors.

\subsubsection{Shower observables}

Showers can be described by a number of key calorimetric observables. The most basic observables are the distributions of the deposited energy within the \grid, representing the longitudinal projection of the grid, as shown in Fig.~\ref{fig:shower_xz}, and the transverse projection, Fig.~\ref{fig:shower_xy}. \textbf{Shower profiles} show how energy is deposited along the longitudinal and radial axis, as depicted for a single shower in Fig.~\ref{fig:grid_profiles}. As we are not comparing single shower profiles, but over a larger statistics, also \textbf{shower profile moments} are validated: the mean (first moment) and variance (second moment). A \textbf{voxel energy distribution} shows the energy distribution within voxels, and spans over many orders of magnitude. The \textbf{number of voxels} shows the number of voxels in which energy was deposited, illustrating the sparsity of the showers.

\subsubsection{General representation}

\begin{figure}[htbp]
    \centering
    \begin{minipage}{0.45\textwidth}
        \centering
    \includegraphics[width=\textwidth,]{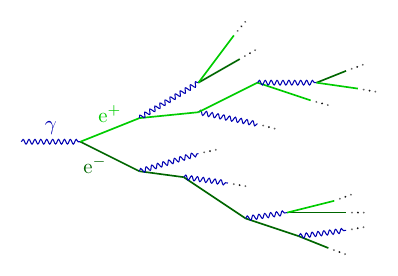}
    \caption{A sketch of the cascade of particles developing in the dense material.}
    \label{fig:cascade}
    \end{minipage}\hfill
    \begin{minipage}{0.45\textwidth}
        \centering
    \includegraphics[width=1\textwidth,trim={1cm 2cm 0 0},clip]{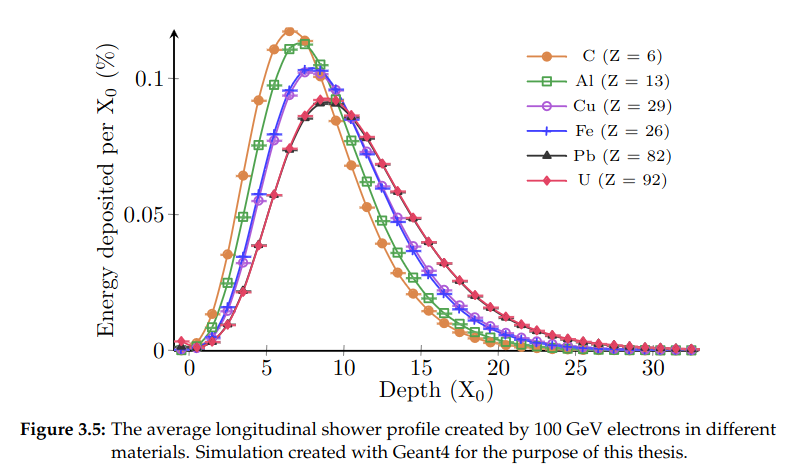}
        \caption{The average longitudinal profile of 100\,GeV electrons in different materials, expressed in units of radiation length $X_0$.}
    \label{fig:thesis_long}
    \end{minipage}
\end{figure}

The amount of the material needed to fully contain the energy of a shower depends on the material of the detector. Showers can be longer or shorter (in the longitudinal direction), and wider or narrower (in the radial direction). A convenient unit to describe showers in the longitudinal direction is the radiation length $X_0$. Radiation length is defined by the amount of material in which high-energetic ($>1$ GeV) particles lose on average $1 -\mathrm{e}^{-1}$ of their energies by bremsstrahlung. Once the longitudinal distance is expressed in units of radiation length, showers developing in different materials become more alike. This is shown in Fig.~\ref{fig:thesis_long}, which shows example longitudinal profiles for showers in different materials. In the radial direction, a unit of Moli\'ere radius $R_M$ is used. On average, 90\% of the initial particle energy is deposited inside the cylinder with radius of size 1.

\subsubsection{Choice of \grid dimensions}

The basic idea behind the \grid is that showers are represented in local cylindrical coordinates, with the size of each voxel corresponding to roughly the same size in units of radiation length and Moli\'ere radius. This allows showers to be represented on a similar scale. The number of the voxels corresponds to the one used in Par04 example of \textsc{Geant4}, or the dataset 2 of CaloChallenge:
$$N\times R\times P = 45 \times 9 \times 16 = 6'480$$
For each detector the size of an individual voxel (expressed in SI units) is different, and is linked to the absorber material, as well as to the structure of the detector. Table~\ref{tab:dim} summarises the dimensions.

\begin{table}[h]
    \centering
    \caption{Dimensions of the chosen \gridfull. There are 45 cells in $z$, 9 cells in $r$ and 16 cells in the $\varphi$ direction. }
    \label{tab:dim}
    \begin{tabular}{lccc}
        \hline
        \textbf{Detector name} & $\mathbf{\Delta r}$\,(mm)& $\mathbf{\Delta}\mathbf{ \varphi}$\,(rad)& $\mathbf{\Delta z}$\,(mm)\\
        \hline
        Par04\_SiW~\cite{ddfastsim_script_par04Si}&4.65&${2\pi}/{16}\approx0.39$&3.4\\
        Par04\_SciPb~\cite{ddfastsim_script_par04SciPb}& 8.0&${2\pi}/{16}$& 5.6 \\
        ODD~\cite{ddfastsim_script_odd}&4.65 &${2\pi}/{16}$&5.05\\
        
        FCCee\_CLD~\cite{ddfastsim_script_cld}& 4.65&${2\pi}/{16}$&5.05\\
        FCCee\_ALLEGRO~\cite{ddfastsim_script_allegro}&5 &${2\pi}/{16}$&11\\
        \hline
    \end{tabular}
\end{table}

\subsection{Incident particles}
\label{sec:values}

\subsubsection{Training datasets}

The \dataset dataset contains electromagnetic showers initiated by incident photons, nearly 1
million per each detector. Some incident particles were removed from the sample to improve sample quality, as they interacted prior to
their entrance to the calorimeter (even though the tracking detectors were removed). For detail numbers see Appendix~\ref{app:numShowers}. The \textbf{energy} of those incident photons ranges from 1\,GeV to 1\,TeV for the Par04 and ODD detectors (designed to mimic hadron collider detectors), and from 1\, GeV to 100\,GeV for FCC-ee detectors, which are designed for electron positron colliders. Unlike most existing datasets, the spectrum of energies is flat, as our studies indicated an improved performance if the training data was not imbalanced.

\begin{figure}
    \centering
    \includegraphics[width=\linewidth]{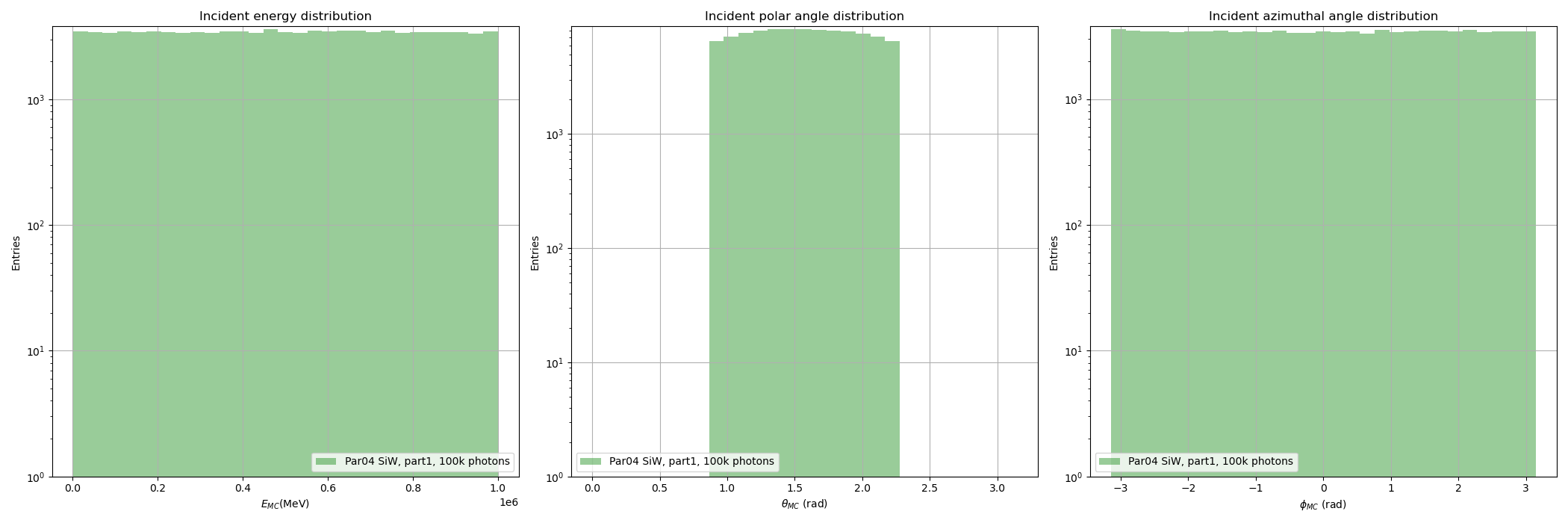}
    \caption{Distributions of incident energy, polar angle, and azimuthal angle for incident particles. Calculated for the first file of the Par04\_SiW main dataset. Plots for all detectors and all files are included in Appendix~\ref{app:training}.}
    \label{fig:plots_training_incident}
\end{figure}

\begin{figure}[htbp]
  \centering
  \begin{subfigure}[t]{0.45\textwidth}
    \includegraphics[width=\textwidth,]{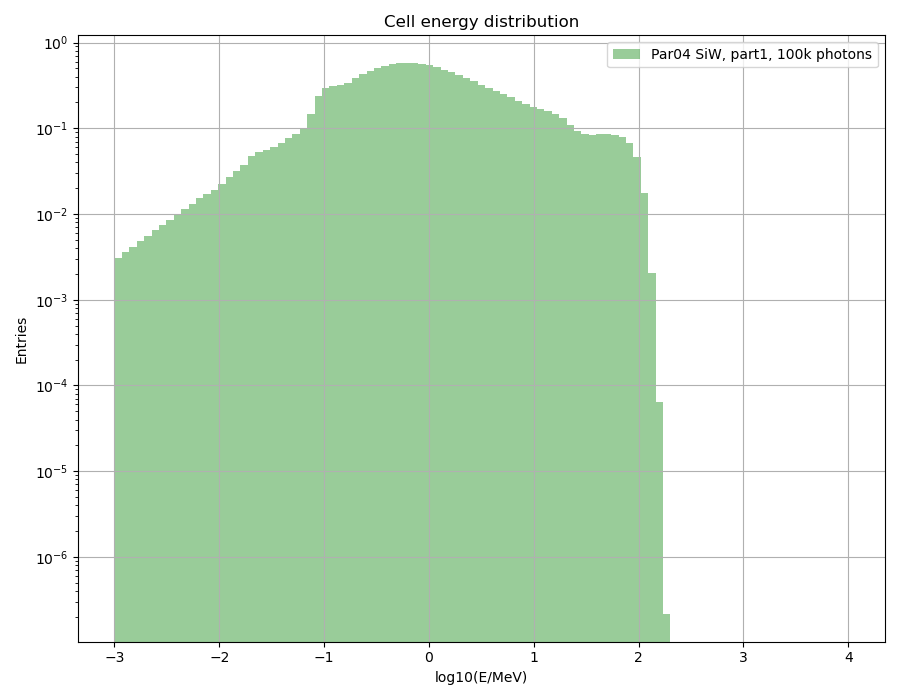}
    \caption{Voxel energy distribution presented as a log10 to highlight the wide range of values.}
    \label{fig:plots_training_logEcell}
  \end{subfigure}
  \hfill
  \begin{subfigure}[t]{0.45\textwidth}
\includegraphics[width=\textwidth]{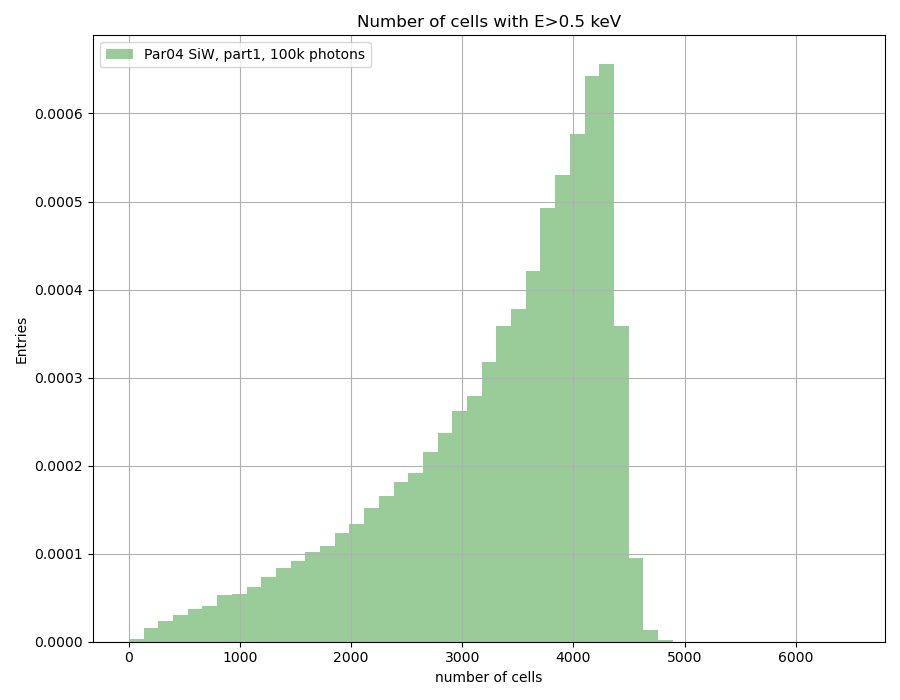}
    \caption{Distribution of number of cells per shower. The width highlights the large difference between showers in the dataset, and shows the sparsity.}
    \label{fig:plots_training_numcells}
  \end{subfigure}
  \caption{Distributions calculated for the first file of the Par04\_SiW main dataset. }
  \label{fig:plots_training}
\end{figure}

The \textbf{geometrical range} in the detector spans from $\theta=0.87$ to $2.27$ (in radians), which corresponds to a pseudorapidity range of $\eta=-0.76$ to $0.76$, meaning only the barrel region was included for all detectors. This is already a much larger detector region than previous datasets, but could certainly be extended to include the endcap and the transition region between barrel and endcap. The coverage in the azimuthal angle is full, from $\phi=-\pi$ to $\pi$. The spectra of incident energies and angles are shown in Fig.~\ref{fig:plots_training_incident}. The main challenges of simulating calorimeter showers are shown in Fig.~\ref{fig:plots_training_logEcell} and Fig.~\ref{fig:plots_training_numcells}, which show the distribution of voxel energy and the number of non-zero voxels, respectively. The wide range of energy values, spanning over several orders of magnitude, and the large number of empty voxels (high sparsity) make these datasets far different to typical RGB pictures. 

\subsubsection{Testing datasets}

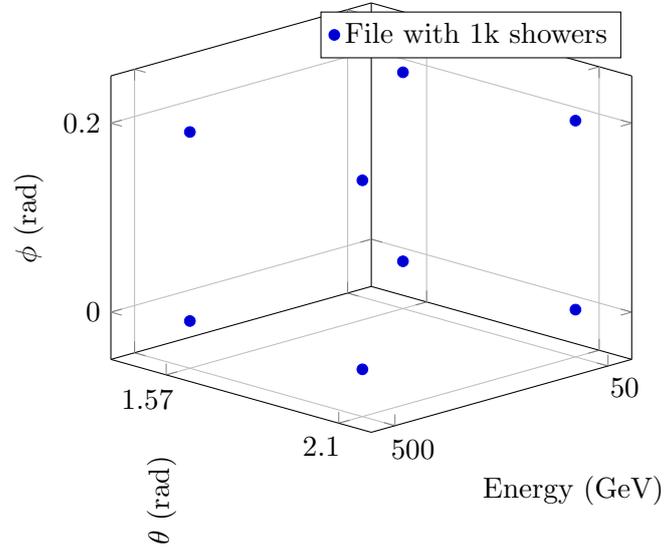
\begin{figure}\centering
\begin{tikzpicture}
\begin{axis}[
    view={135}{20},
    xlabel={Energy (GeV)},
    ylabel={$\theta$ (rad)},
    zlabel={$\phi$ (rad)},
    xmin=0, xmax=550,
    ymin=1.4, ymax=2.2,
    zmin=-0.05, zmax=0.25,
    xtick={50,500},
    ytick={1.57,2.1},
    ztick={0,0.2},
    grid=major,
    minor tick num=0,
    axis lines=box
]

\addplot3+[
    only marks,
    mark=*,
    mark size=2
] coordinates {
    (50, 1.57, 0.0)
    (50, 1.57, 0.2)
    (50, 2.10, 0.0)
    (50, 2.10, 0.2)
    (500, 1.57, 0.0)
    (500, 1.57, 0.2)
    (500, 2.10, 0.0)
    (500, 2.10, 0.2)
};
\legend{File with 1k showers}
\end{axis}
\end{tikzpicture}
\caption{The content of the testing dataset: 1000 showers were simulated on a specific point in the phasespace instead of a uniform distribution. Each point corresponds to one file.}
\label{fig:dataset_grid}
\end{figure}

It is typical in detector studies to assess the performance on specific points in the phasespace. For this reason, a small testing dataset has been produced for each of the detectors, which can be used to produce physics validation observables. 1'000 showers are simulated for each of the points on the grid represented by two energy points: 50 and 500 GeV for Par04 and ODD, two $\theta$ angles: 1.57, and 2.1, and finally two $\phi$ angles: 0.0 and 0.2, as shown in Fig.~\ref{fig:dataset_grid}. For FCCee detectors, energy values of 5 and 50 GeV are checked.

\begin{figure}[htbp]
  \centering
  \begin{subfigure}[t]{0.45\textwidth}
    \includegraphics[width=\textwidth,]{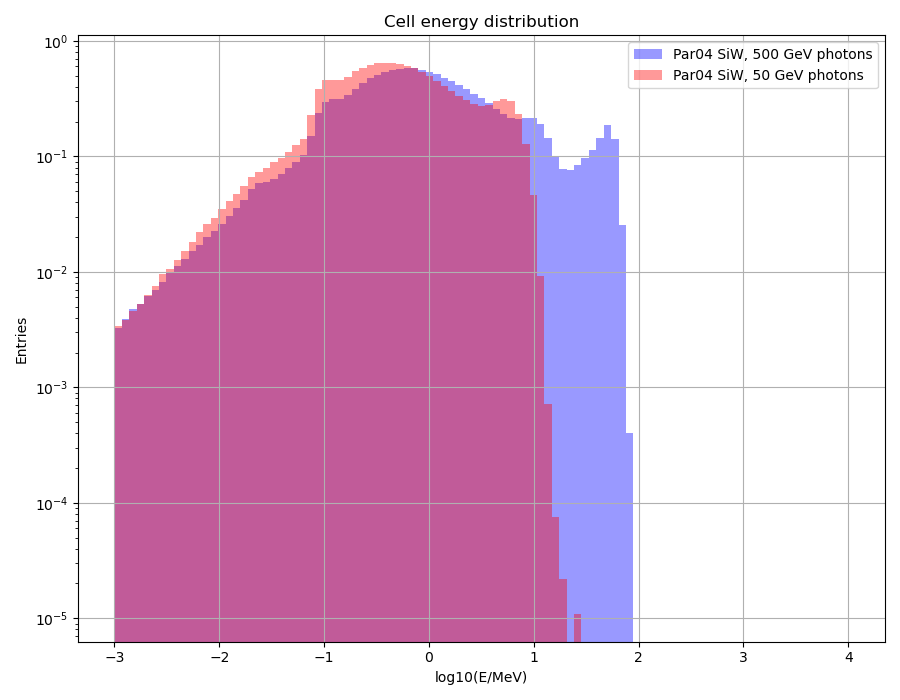}
    \caption{Voxel energy distribution presented with a log10 scale to highlight the wide range of values.}
    \label{fig:plots_validation_logEcell}
  \end{subfigure}
  \hfill
  \begin{subfigure}[t]{0.45\textwidth}
\includegraphics[width=\textwidth]{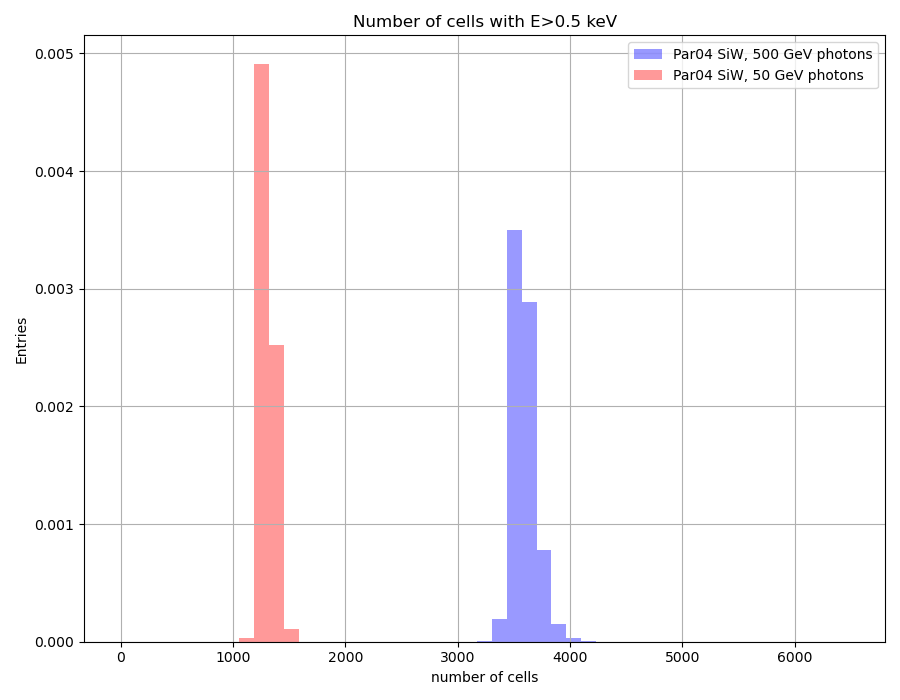}
    \caption{Distribution of the number of cells per shower. The width highlights the large difference between showers in the dataset, and shows the sparsity.}
    \label{fig:plots_validation_numcells}
  \end{subfigure}
  \caption{Distributions calculated for the Par04\_SiW validation datasets at $\theta=\pi$ and $\phi=0$. Plots for all detectors and all testing files are included in Appendix~\ref{app:testing}.}
\end{figure}

\begin{figure}
    \centering
    \includegraphics[width=\linewidth]{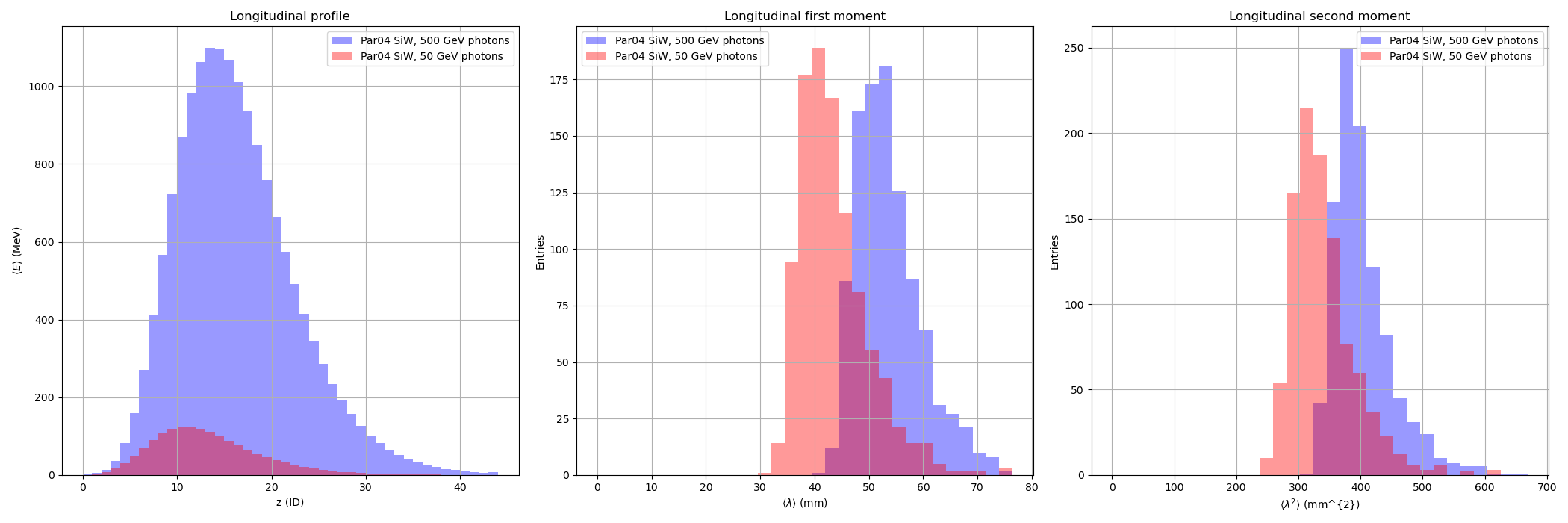}
    \caption{Longitudinal profile and shower moments calculated for the validation files of the Par04\_SiW dataset at $\theta=\pi$ and $\phi=0$.}
    \label{fig:plots_validation_long}
\end{figure}

\begin{figure}
    \centering
    \includegraphics[width=\linewidth]{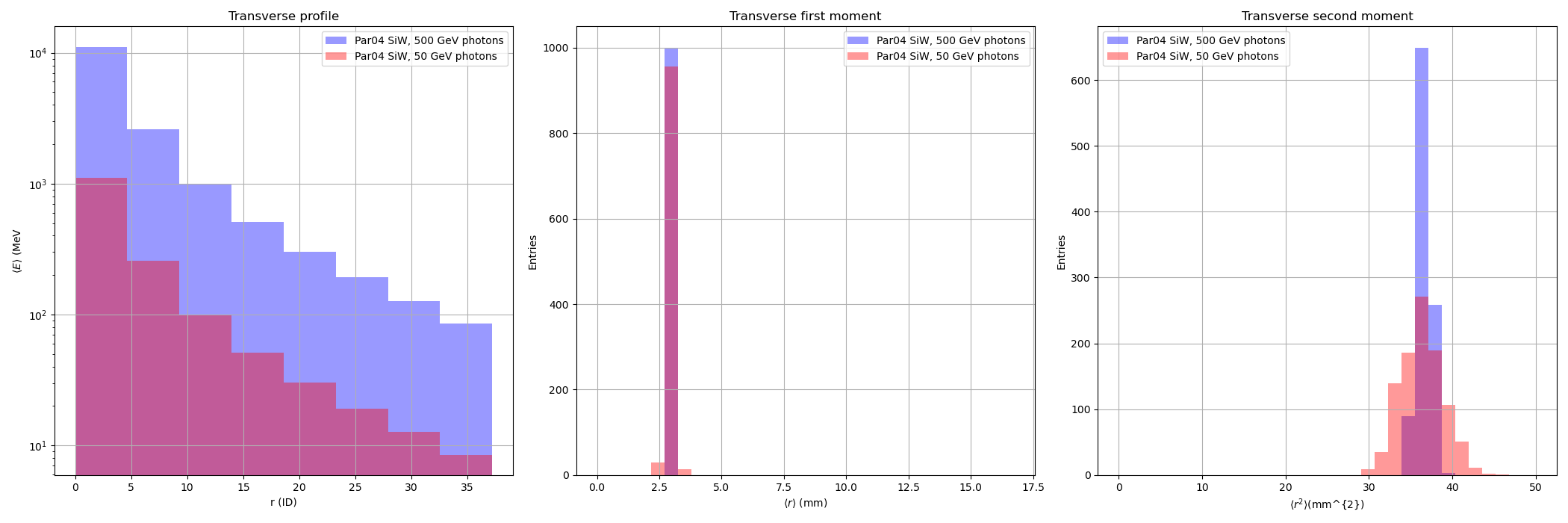}
    \caption{Transverse profile and shower moments calculated for the validation files of the Par04\_SiW dataset at $\theta=\pi$ and $\phi=0$.}
    \label{fig:plots_validation_trans}
\end{figure}

\begin{figure}[htbp]
  \centering
  \begin{subfigure}[b]{0.45\textwidth}
    \includegraphics[width=\textwidth,]{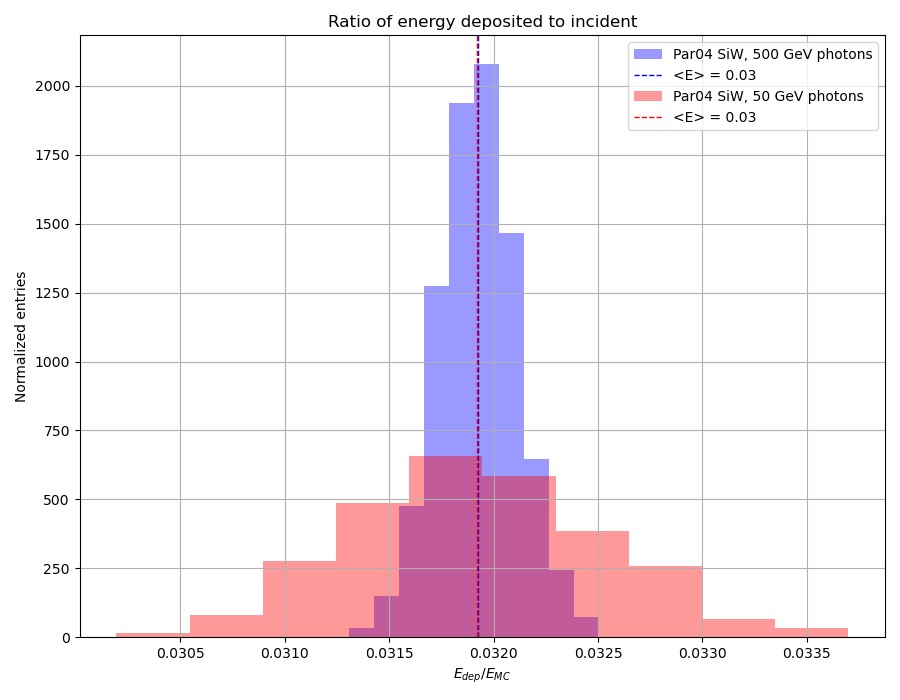}
    \caption{Par04 SiW detector.}
    \label{fig:plots_validation_par04_eratio}
  \end{subfigure}
  \hfill
  \begin{subfigure}[b]{0.45\textwidth}
\includegraphics[width=\textwidth]{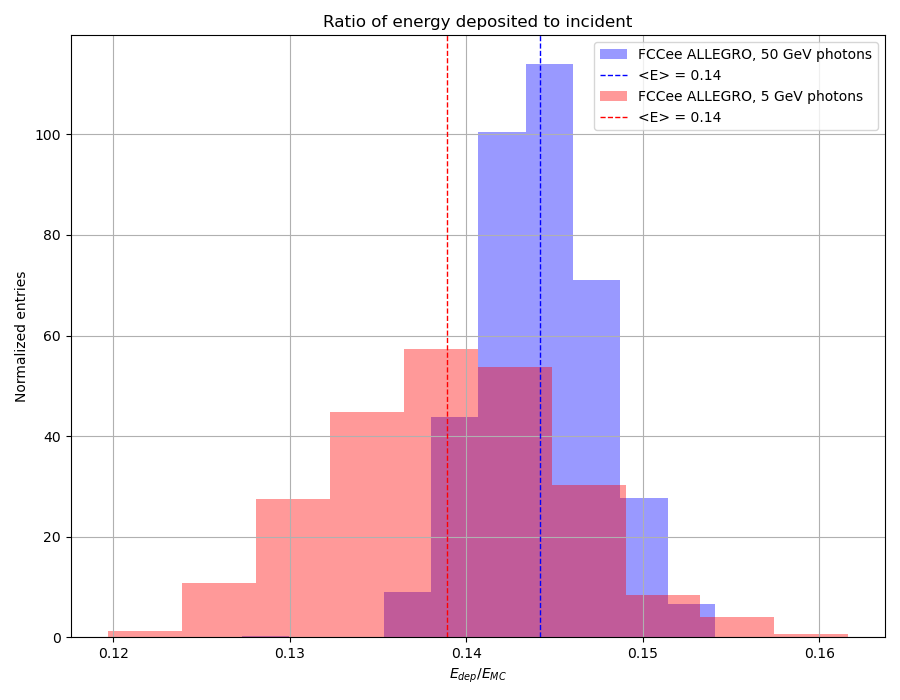}
    \caption{FCCee ALLEGRO detector.}
    \label{fig:plots_validation_allegro_eratio}
  \end{subfigure}
  \caption{Distribution of energy ratio (total energy deposited in the detector to incident particle energy) calculated for testing datasets at $\theta=\pi$ and $\phi=0$. It represents the fraction of energy that is deposited in the sensitive material of the sampling calorimeter.}
\end{figure}

The plots for the validation dataset clearly show the differences with the energy of incident particles (smaller differences appear for different angles). The logarithm of the voxel energy distribution is shown in Fig~\ref{fig:plots_validation_logEcell}, the sparsity of the showers in Fig.~\ref{fig:plots_validation_numcells}, and the longitudinal and transverse profiles and moments in Fig.~\ref{fig:plots_validation_long} and \ref{fig:plots_validation_trans}, respectively. The final plot of the Par04\_SiW geometry represents the distribution of the ratio of energy deposited in the calorimeter to the incident particle energy, Fig.~\ref{fig:plots_validation_par04_eratio}. As can be seen, the ratio is not close to value of 1, as it is a sampling calorimeter, meaning that only a fraction of energy is deposited in the sensitive material. The same can be seen in Fig.~\ref{fig:plots_validation_allegro_eratio} for the FCCee\_ALLEGRO calorimeter, although the ratio is centred around a different values, specific for this calorimeter. Table ~\ref{tab:sf} summarises the mean values calculated for the training datasets (averaged over energies). Those values could be used to scale up the deposited energy, which is typically done during the later stages of data processing. For instance, scaling up with the fraction of energy (known as the sampling fraction) was performed for the CaloChallenge datasets~\cite{faucci_giannelli_2022_6366271_ds2,faucci_giannelli_2022_6366324_ds3}. It may not be required, though, by the experiments which aim at replacing the simulation only (with all calibration coming later, including for full/detailed and fast simulation).
\label{sec:sampling_fraction}
\begin{table}[h]
    \centering
    \caption{Mean sampling fractions for the training datasets (averaged over all incident energies and angles). }
    \label{tab:sf}
    \begin{tabular}{lc}
        \hline
        \textbf{Detector name} & \textbf{Mean sampling fraction} \\
        \hline
        Par04\_SiW&  0.0321\\
        Par04\_SciPb& 0.0330 \\
        ODD& 0.0255\\
        FCCee\_CLD& 0.0257 \\
        FCCee\_ALLEGRO& 0.143 \\
        \hline
    \end{tabular}
\end{table}


\section{Data Records}

The \dataset dataset is released on Zenodo~\cite{lemurs}, with both the training dataset and the testing dataset. 

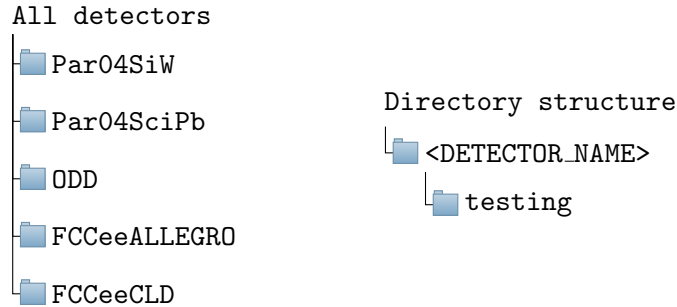
\begin{figure}[ht]
    \centering
\begin{minipage}{0.3\textwidth}
\begin{forest}
  for tree={font=\ttfamily,grow'=0,child anchor=west,parent anchor=south, anchor=west,calign=first,inner xsep=7pt, edge path={\noexpand\path [draw, \forestoption{edge}](!u.south west) +(7.5pt,0) |- (.child anchor) pic {folder} \forestoption{edge label};}, before typesetting nodes={
      if n=1
        {insert before={[,phantom]}}
        {}
    }, fit=band,before computing xy={l=15pt},
  }  
[All detectors
  [Par04SiW
  ]
  [Par04SciPb
  ]
  [ODD
  ]
  [FCCeeALLEGRO
  ]
  [FCCeeCLD
  ]
]
\end{forest}
\end{minipage}
\begin{minipage}{0.3\textwidth}
\begin{forest}
  for tree={font=\ttfamily,grow'=0,child anchor=west,parent anchor=south, anchor=west,calign=first,inner xsep=7pt, edge path={\noexpand\path [draw, \forestoption{edge}](!u.south west) +(7.5pt,0) |- (.child anchor) pic {folder} \forestoption{edge label};}, before typesetting nodes={
      if n=1
        {insert before={[,phantom]}}
        {}
    }, fit=band,before computing xy={l=15pt},
  }  
[Directory structure
  [<DETECTOR\_NAME>
    [testing
    ]
  ]
]
\end{forest}
\end{minipage}
    \caption{Structure of the archive that contains the testing dataset. Files for each detector can be found in the directory with the name of the detector, and then \texttt{testing/} subdirectory.}
    \label{fig:directory}
\end{figure}

The testing dataset is presented in a structure that allows easy navigation between different detectors, as depicted in Fig.~\ref{fig:directory}. The names of the HDF5 (\texttt{.h5}) files included in those directories correspond to the ranges of incident particle observables as described in Sec.~\ref{sec:values}.

The training files are all uploaded separately (as Zenodo does not support directory structure) to allow users to choose how many of those large files are useful for their studies. They can be easily recognized by the name of the file: \\\texttt{LEMURS\_<DETECTOR>\_gamma\_100kEvents\_1GeV<MAX>\_GPSflat\_part<PART>.h5}.

The structure of all files is inspired by the structure of CaloChallenge files to minimise any modifications needed to reuse its tools. It must be noted it is not identical and the biggest change is the representation of a single shower. The HDF5 structure is described in Table~\ref{tab:h5}. $S$ is the number of showers in the file, $N=45$ the number of layers, $R=9$ the number of radial bins, and $P=16$ the number of longitudinal bins.

\begin{table}[h]
    \centering
    \caption{Structure of the HDF5 dataset used for training. All datasets are stored in \texttt{float32} precision and compressed with gzip (level~9).}
    \label{tab:h5}
    \begin{tabular}{lll}
        \hline
        \textbf{Dataset name} & \textbf{Shape} & \textbf{Description} \\
        \hline
        \texttt{incident\_energy} & $(S)$          & Incident particle energy ($E_\mathrm{MC}$) \\
        \texttt{incident\_phi}    & $(S)$          & Incident azimuthal angle ($\phi_\mathrm{MC}$) \\
        \texttt{incident\_theta}  & $(S)$          & Incident polar angle ($\theta_\mathrm{MC}$) \\
        \texttt{showers}          & $(S, N, R, P)$ & Showers in \grid \\
        \hline
    \end{tabular}
\end{table}

\section{Technical Validation}
Validation presented in this paper has been done using the validation script of the \texttt{ddfastsim} package~\cite{ddfastsim_script_validation}. Figures \ref{fig:plots_training_incident} and \ref{fig:plots_training} are done for the first file of the main dataset for the Par04\_SiW detector, with all incident energies and angles being represented. To see the differences in the distributions for different incident energies, the smaller validation datasets were used, as shown in Figures~\ref{fig:plots_validation_logEcell} to \ref{fig:plots_validation_allegro_eratio}), for Par04\_SiW and FCCee\_ALLEGRO detectors.

The ultimate validation of LEMURS has been demonstrated in the pre-training of the CaloDiT-2 model released in the standard simulation toolkit Geant4~\cite{Par04}.

\section{Usage Notes}
This dataset can be used either picking up individual detectors, or combining all to build a generalisable model. The main datasets contain the continuous spectra of incident conditions (energy, polar angle, azimuthal angle), while validation datasets are meant for final validation of physics variables that are much easier to assess with a substantial sample simulated at specific incident conditions.

\section{Code Availability}
This dataset has been generated with ddfastsim repo~\cite{ddfastsim}. The software used for generation is key4hep~\cite{Carceller:2025ydg}, which in particular includes the definitions of FCC-ee detectors~\cite{k4geo}. The Open Data Detector repository can be accessed from~\cite{odd}.

All configuration files used to steer the simulation and define the dimensions, incident particles (for both the training and testing datasets) are included in \texttt{ddfastsim} in \texttt{options/<DETECTOR\_NAME>}. The scripts that were used on HTCondor for wide-scale production are placed in the \texttt{condor} directory. To be reused they would require well-defined user-specific and cluster-specific changes. The plotting (validation) scripts are placed under the \texttt{scripts} directory.

The original implementation of the \grid and the event display can be found in Geant4 Par04 extended example, with the most recent changes included in the 11.4.beta release~\cite{Par04}.

\section{Data Availability}
The \dataset dataset is released on Zenodo~\cite{lemurs}. As it surpasses the quota on the total size, a special permission was granted to store all 61 GB together.

\section{Funding}
This work benefited from support by the CERN Strategic R\&D Programme on Technologies for
Future Experiments and has received funding from the European Union’s Horizon 2020 Research
and Innovation programme under Grant Agreement No. 101004761.

\bibliographystyle{unsrt_custom}
\bibliography{references}

\appendix
\section{Data multiplicity}
\label{app:numShowers}

Table~\ref{tab:numShowers} summarizes the training and validation datasets used in this work.  
For each detector configuration, we list the file name and the number of events contained in the file.  

\begin{longtable}{lr}
\caption{List of dataset files and number of events.\label{tab:numShowers}}\\
\hline
\textbf{File} & \textbf{\# Events} \\
\hline
\endfirsthead
\hline
\textbf{File} & \textbf{\# Events} \\
\hline
\endhead
\hline
\endfoot

LEMURS\_Par04SiW\_gamma\_100kEvents\_1GeV1TeV\_GPSflat\_part1.h5 & 100000 \\
LEMURS\_Par04SiW\_gamma\_100kEvents\_1GeV1TeV\_GPSflat\_part10.h5 & 100000 \\
LEMURS\_Par04SiW\_gamma\_100kEvents\_1GeV1TeV\_GPSflat\_part2.h5 & 100000 \\
LEMURS\_Par04SiW\_gamma\_100kEvents\_1GeV1TeV\_GPSflat\_part3.h5 & 100000 \\
LEMURS\_Par04SiW\_gamma\_100kEvents\_1GeV1TeV\_GPSflat\_part4.h5 & 100000 \\
LEMURS\_Par04SiW\_gamma\_100kEvents\_1GeV1TeV\_GPSflat\_part5.h5 & 100000 \\
LEMURS\_Par04SiW\_gamma\_100kEvents\_1GeV1TeV\_GPSflat\_part6.h5 & 100000 \\
LEMURS\_Par04SiW\_gamma\_100kEvents\_1GeV1TeV\_GPSflat\_part7.h5 & 100000 \\
LEMURS\_Par04SiW\_gamma\_100kEvents\_1GeV1TeV\_GPSflat\_part8.h5 & 100000 \\
LEMURS\_Par04SiW\_gamma\_100kEvents\_1GeV1TeV\_GPSflat\_part9.h5 & 100000 \\
LEMURS\_Par04SiW\_gamma\_1000events\_500GeV\_phi0.0\_theta1.57.h5 & 1000 \\
LEMURS\_Par04SiW\_gamma\_1000events\_500GeV\_phi0.0\_theta2.1.h5 & 1000 \\
LEMURS\_Par04SiW\_gamma\_1000events\_500GeV\_phi0.2\_theta1.57.h5 & 1000 \\
LEMURS\_Par04SiW\_gamma\_1000events\_500GeV\_phi0.2\_theta2.1.h5 & 1000 \\
LEMURS\_Par04SiW\_gamma\_1000events\_50GeV\_phi0.0\_theta1.57.h5 & 1000 \\
LEMURS\_Par04SiW\_gamma\_1000events\_50GeV\_phi0.0\_theta2.1.h5 & 1000 \\
LEMURS\_Par04SiW\_gamma\_1000events\_50GeV\_phi0.2\_theta1.57.h5 & 1000 \\
LEMURS\_Par04SiW\_gamma\_1000events\_50GeV\_phi0.2\_theta2.1.h5 & 1000 \\
LEMURS\_Par04SciPb\_gamma\_100kEvents\_1GeV1TeV\_GPSflat\_part1.h5 & 100000 \\
LEMURS\_Par04SciPb\_gamma\_100kEvents\_1GeV1TeV\_GPSflat\_part10.h5 & 100000 \\
LEMURS\_Par04SciPb\_gamma\_100kEvents\_1GeV1TeV\_GPSflat\_part2.h5 & 100000 \\
LEMURS\_Par04SciPb\_gamma\_100kEvents\_1GeV1TeV\_GPSflat\_part3.h5 & 100000 \\
LEMURS\_Par04SciPb\_gamma\_100kEvents\_1GeV1TeV\_GPSflat\_part4.h5 & 100000 \\
LEMURS\_Par04SciPb\_gamma\_100kEvents\_1GeV1TeV\_GPSflat\_part5.h5 & 100000 \\
LEMURS\_Par04SciPb\_gamma\_100kEvents\_1GeV1TeV\_GPSflat\_part6.h5 & 100000 \\
LEMURS\_Par04SciPb\_gamma\_100kEvents\_1GeV1TeV\_GPSflat\_part7.h5 & 100000 \\
LEMURS\_Par04SciPb\_gamma\_100kEvents\_1GeV1TeV\_GPSflat\_part8.h5 & 100000 \\
LEMURS\_Par04SciPb\_gamma\_100kEvents\_1GeV1TeV\_GPSflat\_part9.h5 & 100000 \\
LEMURS\_Par04SciPb\_gamma\_1000events\_500GeV\_phi0.0\_theta1.57.h5 & 1000 \\
LEMURS\_Par04SciPb\_gamma\_1000events\_500GeV\_phi0.0\_theta2.1.h5 & 1000 \\
LEMURS\_Par04SciPb\_gamma\_1000events\_500GeV\_phi0.2\_theta1.57.h5 & 1000 \\
LEMURS\_Par04SciPb\_gamma\_1000events\_500GeV\_phi0.2\_theta2.1.h5 & 1000 \\
LEMURS\_Par04SciPb\_gamma\_1000events\_50GeV\_phi0.0\_theta1.57.h5 & 1000 \\
LEMURS\_Par04SciPb\_gamma\_1000events\_50GeV\_phi0.0\_theta2.1.h5 & 1000 \\
LEMURS\_Par04SciPb\_gamma\_1000events\_50GeV\_phi0.2\_theta1.57.h5 & 1000 \\
LEMURS\_Par04SciPb\_gamma\_1000events\_50GeV\_phi0.2\_theta2.1.h5 & 1000 \\
LEMURS\_ODD\_gamma\_100kEvents\_1GeV1TeV\_GPSflat\_part1.h5 & 100000 \\
LEMURS\_ODD\_gamma\_100kEvents\_1GeV1TeV\_GPSflat\_part10.h5 & 100000 \\
LEMURS\_ODD\_gamma\_100kEvents\_1GeV1TeV\_GPSflat\_part2.h5 & 100000 \\
LEMURS\_ODD\_gamma\_100kEvents\_1GeV1TeV\_GPSflat\_part3.h5 & 100000 \\
LEMURS\_ODD\_gamma\_100kEvents\_1GeV1TeV\_GPSflat\_part4.h5 & 100000 \\
LEMURS\_ODD\_gamma\_100kEvents\_1GeV1TeV\_GPSflat\_part5.h5 & 99500 \\
LEMURS\_ODD\_gamma\_100kEvents\_1GeV1TeV\_GPSflat\_part6.h5 & 100000 \\
LEMURS\_ODD\_gamma\_100kEvents\_1GeV1TeV\_GPSflat\_part7.h5 & 100000 \\
LEMURS\_ODD\_gamma\_100kEvents\_1GeV1TeV\_GPSflat\_part8.h5 & 100000 \\
LEMURS\_ODD\_gamma\_100kEvents\_1GeV1TeV\_GPSflat\_part9.h5 & 100000 \\
LEMURS\_ODD\_gamma\_1000events\_500GeV\_phi0.0\_theta1.57.h5 & 1000 \\
LEMURS\_ODD\_gamma\_1000events\_500GeV\_phi0.0\_theta2.1.h5 & 1000 \\
LEMURS\_ODD\_gamma\_1000events\_500GeV\_phi0.2\_theta1.57.h5 & 1000 \\
LEMURS\_ODD\_gamma\_1000events\_500GeV\_phi0.2\_theta2.1.h5 & 1000 \\
LEMURS\_ODD\_gamma\_1000events\_50GeV\_phi0.0\_theta1.57.h5 & 1000 \\
LEMURS\_ODD\_gamma\_1000events\_50GeV\_phi0.0\_theta2.1.h5 & 1000 \\
LEMURS\_ODD\_gamma\_1000events\_50GeV\_phi0.2\_theta1.57.h5 & 1000 \\
LEMURS\_ODD\_gamma\_1000events\_50GeV\_phi0.2\_theta2.1.h5 & 1000 \\
LEMURS\_FCCeeCLD\_gamma\_100kEvents\_1GeV100GeV\_GPSflat\_part1.h5 & 100000 \\
LEMURS\_FCCeeCLD\_gamma\_100kEvents\_1GeV100GeV\_GPSflat\_part10.h5 & 89846 \\
LEMURS\_FCCeeCLD\_gamma\_100kEvents\_1GeV100GeV\_GPSflat\_part2.h5 & 100000 \\
LEMURS\_FCCeeCLD\_gamma\_100kEvents\_1GeV100GeV\_GPSflat\_part3.h5 & 100000 \\
LEMURS\_FCCeeCLD\_gamma\_100kEvents\_1GeV100GeV\_GPSflat\_part4.h5 & 100000 \\
LEMURS\_FCCeeCLD\_gamma\_100kEvents\_1GeV100GeV\_GPSflat\_part5.h5 & 89900 \\
LEMURS\_FCCeeCLD\_gamma\_100kEvents\_1GeV100GeV\_GPSflat\_part6.h5 & 89812 \\
LEMURS\_FCCeeCLD\_gamma\_100kEvents\_1GeV100GeV\_GPSflat\_part7.h5 & 89880 \\
LEMURS\_FCCeeCLD\_gamma\_100kEvents\_1GeV100GeV\_GPSflat\_part8.h5 & 89911 \\
LEMURS\_FCCeeCLD\_gamma\_100kEvents\_1GeV100GeV\_GPSflat\_part9.h5 & 89738 \\
LEMURS\_FCCeeCLD\_gamma\_1000events\_50GeV\_phi0.0\_theta1.57.h5 & 926 \\
LEMURS\_FCCeeCLD\_gamma\_1000events\_50GeV\_phi0.0\_theta2.1.h5 & 887 \\
LEMURS\_FCCeeCLD\_gamma\_1000events\_50GeV\_phi0.2\_theta1.57.h5 & 912 \\
LEMURS\_FCCeeCLD\_gamma\_1000events\_50GeV\_phi0.2\_theta2.1.h5 & 887 \\
LEMURS\_FCCeeCLD\_gamma\_1000events\_5GeV\_phi0.0\_theta1.57.h5 & 988 \\
LEMURS\_FCCeeCLD\_gamma\_1000events\_5GeV\_phi0.0\_theta2.1.h5 & 977 \\
LEMURS\_FCCeeCLD\_gamma\_1000events\_5GeV\_phi0.2\_theta1.57.h5 & 982 \\
LEMURS\_FCCeeCLD\_gamma\_1000events\_5GeV\_phi0.2\_theta2.1.h5 & 986 \\
LEMURS\_FCCeeALLEGRO\_gamma\_100kEvents\_1GeV100GeV\_GPSflat\_part1.h5 & 99942 \\
LEMURS\_FCCeeALLEGRO\_gamma\_100kEvents\_1GeV100GeV\_GPSflat\_part10.h5 & 99942 \\
LEMURS\_FCCeeALLEGRO\_gamma\_100kEvents\_1GeV100GeV\_GPSflat\_part2.h5 & 99939 \\
LEMURS\_FCCeeALLEGRO\_gamma\_100kEvents\_1GeV100GeV\_GPSflat\_part3.h5 & 99923 \\
LEMURS\_FCCeeALLEGRO\_gamma\_100kEvents\_1GeV100GeV\_GPSflat\_part4.h5 & 99931 \\
LEMURS\_FCCeeALLEGRO\_gamma\_100kEvents\_1GeV100GeV\_GPSflat\_part5.h5 & 99925 \\
LEMURS\_FCCeeALLEGRO\_gamma\_100kEvents\_1GeV100GeV\_GPSflat\_part6.h5 & 99935 \\
LEMURS\_FCCeeALLEGRO\_gamma\_100kEvents\_1GeV100GeV\_GPSflat\_part7.h5 & 99915 \\
LEMURS\_FCCeeALLEGRO\_gamma\_100kEvents\_1GeV100GeV\_GPSflat\_part8.h5 & 99935 \\
LEMURS\_FCCeeALLEGRO\_gamma\_100kEvents\_1GeV100GeV\_GPSflat\_part9.h5 & 99923 \\
LEMURS\_FCCeeALLEGRO\_gamma\_1000events\_50GeV\_phi0.0\_theta1.57.h5 & 998 \\
LEMURS\_FCCeeALLEGRO\_gamma\_1000events\_50GeV\_phi0.0\_theta2.1.h5 & 999 \\
LEMURS\_FCCeeALLEGRO\_gamma\_1000events\_50GeV\_phi0.2\_theta1.57.h5 & 999 \\
LEMURS\_FCCeeALLEGRO\_gamma\_1000events\_50GeV\_phi0.2\_theta2.1.h5 & 999 \\
LEMURS\_FCCeeALLEGRO\_gamma\_1000events\_5GeV\_phi0.0\_theta1.57.h5 & 1000 \\
LEMURS\_FCCeeALLEGRO\_gamma\_1000events\_5GeV\_phi0.0\_theta2.1.h5 & 1000 \\
LEMURS\_FCCeeALLEGRO\_gamma\_1000events\_5GeV\_phi0.2\_theta1.57.h5 & 1000 \\
LEMURS\_FCCeeALLEGRO\_gamma\_1000events\_5GeV\_phi0.2\_theta2.1.h5 & 1000 \\

\end{longtable}

\section{Validation plots of training dataset}
\label{app:training}

Plots for each of the training files are presented here for completeness of validation. Only incident energy and angles are included in this appendix. The drawing script used to produce the files is in \texttt{ddfastsim}~\cite{ddfastsim} repository: \texttt{zenodo\_utils/run\_draws.sh}.

\foreach \i in {2,3,4,5,6,7,8,9,10}{%
\begin{figure}[!htb]
    \centering
    \includegraphics[width=\linewidth]{images/Par04SiW_part\i_incident_distr.png}
    \caption{Distributions of incident energy, polar angle, and azimuthal angle. Calculated for \texttt{LEMURS\_Par04SiW\_gamma\_100kEvents\_1GeV1TeV\_GPSflat\_part\i.h5}.}
\end{figure}%
}

\foreach \i in {1,2,3,4,5,6,7,8,9,10}{%
\begin{figure}[!htb]
    \centering
    \includegraphics[width=\linewidth]{images/Par04SciPb_part\i_incident_distr.png}
    \caption{Distributions of incident energy, polar angle, and azimuthal angle. Calculated for \texttt{LEMURS\_Par04SciPb\_gamma\_100kEvents\_1GeV1TeV\_GPSflat\_part\i.h5}.}
\end{figure}%
}

\foreach \i in {1,2,3,4,5,6,7,8,9,10}{%
\begin{figure}[!htb]
    \centering
    \includegraphics[width=\linewidth]{images/ODD_part\i_incident_distr.png}
    \caption{Distributions of incident energy, polar angle, and azimuthal angle. Calculated for \texttt{LEMURS\_ODD\_gamma\_100kEvents\_1GeV1TeV\_GPSflat\_part\i.h5}.}
\end{figure}%
}

\foreach \i in {1,2,3,4,5,6,7,8,9,10}{%
\begin{figure}[!htb]
    \centering
    \includegraphics[width=\linewidth]{images/FCCeeCLD_part\i_incident_distr.png}
    \caption{Distributions of incident energy, polar angle, and azimuthal angle. Calculated for \texttt{LEMURS\_FCCeeCLD\_gamma\_100kEvents\_1GeV100GeV\_GPSflat\_part\i.h5}.}
\end{figure}%
}
\foreach \i in {1,2,3,4,5,6,7,8,9,10}{%
\begin{figure}[!htb]
    \centering
    \includegraphics[width=\linewidth]{images/FCCeeALLEGRO_part\i_incident_distr.png}
    \caption{Distributions of incident energy, polar angle, and azimuthal angle. Calculated for \texttt{LEMURS\_FCCeeALLEGRO\_gamma\_100kEvents\_1GeV100GeV\_GPSflat\_part\i.h5}.}
\end{figure}%
}

\clearpage
\section{Validation plots of testing dataset}
\label{app:testing}

Plots for each of the testing files are presented here for completeness of validation. Only voxel energy spectrum and number of non-empty voxels are included in this appendix. The drawing script used to produce the files is in \texttt{ddfastsim}~\cite{ddfastsim} repository: \texttt{zenodo\_utils/run\_draws.sh}. More observables can be plotted (as shown in the main part of this note).

\foreach \det in {Par04SiW, Par04SciPb,ODD}{%
\foreach \myphi in {0.0, 0.2}{%
\foreach \myth in {1.57, 2.1}{%
\begin{figure}[htbp]
  \centering
  \begin{subfigure}[t]{0.45\textwidth}
    \includegraphics[width=\textwidth,]{images/\det_50vs500GeV_phi\myphi_theta\myth.cell_energy_log.png}
    \caption{Voxel energy distribution presented with a log10 scale.}
  \end{subfigure}
  \hfill
  \begin{subfigure}[t]{0.45\textwidth}
\includegraphics[width=\textwidth]{images/\det_50vs500GeV_phi\myphi_theta\myth.num_cells.png}
    \caption{Distribution of the number of voxels per shower.}
  \end{subfigure}
  \caption{Distributions calculated for \texttt{LEMURS\_\det\_gamma\_1000events\_50GeV\_phi\myphi\_theta\myth.h5} and \texttt{LEMURS\_\det\_gamma\_1000events\_500GeV\_phi\myphi\_theta\myth.h5}.}
\end{figure}%
}}}

\foreach \det in {FCCeeCLD,FCCeeALLEGRO}{%
\foreach \myphi in {0.0, 0.2}{%
\foreach \myth in {1.57, 2.1}{%
\begin{figure}[htbp]
  \centering
  \begin{subfigure}[t]{0.45\textwidth}
    \includegraphics[width=\textwidth,]{images/\det_5vs50GeV_phi\myphi_theta\myth.cell_energy_log.png}
    \caption{Voxel energy distribution presented with a log10 scale.}
  \end{subfigure}
  \hfill
  \begin{subfigure}[t]{0.45\textwidth}
\includegraphics[width=\textwidth]{images/\det_5vs50GeV_phi\myphi_theta\myth.num_cells.png}
    \caption{Distribution of the number of voxels per shower.}
  \end{subfigure}
  \caption{Distributions calculated for \texttt{LEMURS\_\det\_gamma\_1000events\_5GeV\_phi\myphi\_theta\myth.h5} and \texttt{LEMURS\_\det\_gamma\_1000events\_50GeV\_phi\myphi\_theta\myth.h5}.}
\end{figure}%
}}}

\end{document}